\documentclass[aps,prb,twocolumn,superscriptaddress,showpacs,showkeys,amsmath,amssymb]{revtex4}

\usepackage{amsfonts}
\usepackage{amssymb,amsmath}
\usepackage{graphicx}
\usepackage{dcolumn}
\usepackage{bm}
\usepackage{color}

\begin{document}

\title{Frustrated quantum Heisenberg antiferromagnets at high magnetic fields:\\
       Beyond the flat-band scenario}

\author{Oleg Derzhko}
\affiliation{Institute for Condensed Matter Physics,
          National Academy of Sciences of Ukraine,
          1 Svientsitskii Street, L'viv-11, 79011, Ukraine}
\affiliation{Department for Theoretical Physics, 
          Ivan Franko National University of L'viv, 
          12 Drahomanov Street, L'viv-5, 79005, Ukraine}
\affiliation{Institut f\"{u}r theoretische Physik,
          Otto-von-Guericke-Universit\"{a}t Magdeburg,
          P.O. Box 4120, D-39016 Magdeburg, Germany}
\affiliation{Abdus Salam International Centre for Theoretical Physics,
          Strada Costiera 11, I-34151 Trieste, Italy}

\author{Johannes Richter}
\affiliation{Institut f\"{u}r theoretische Physik,
          Otto-von-Guericke-Universit\"{a}t Magdeburg,
          P.O. Box 4120, D-39016 Magdeburg, Germany}

\author{Olesia Krupnitska}
\affiliation{Department for Theoretical Physics, 
          Ivan Franko National University of L'viv, 
          12 Drahomanov Street, L'viv-5, 79005, Ukraine}

\author{Taras Krokhmalskii}
\affiliation{Institute for Condensed Matter Physics,
          National Academy of Sciences of Ukraine,
          1 Svientsitskii Street, L'viv-11, 79011, Ukraine}
\affiliation{Department for Theoretical Physics, 
          Ivan Franko National University of L'viv, 
          12 Drahomanov Street, L'viv-5, 79005, Ukraine}

\date{\today}



\begin{abstract}
We consider the spin-1/2 antiferromagnetic Heisenberg model on three frustrated lattices
(the diamond chain, the dimer-plaquette chain and the two-dimensional square-kagome lattice)
with almost dispersionless lowest magnon band.
Eliminating high-energy degrees of freedom at high magnetic fields,
we construct low-energy effective Hamiltonians 
which are much simpler than the initial ones. 
These  effective Hamiltonians allow a more extended analytical and numerical analysis.
In addition to the standard strong-coupling perturbation theory 
we also use a localized-magnon based approach leading to a substantial improvement of the strong-coupling approximation.
We perform extensive exact diagonalization calculations 
to check the quality of different effective Hamiltonians by comparison with the initial models.
Based on the effective-model description
we examine the low-temperature properties of the considered frustrated quantum Heisenberg antiferromagnets in the high-field regime.
We also apply our approach to explore thermodynamic properties for a generalized diamond spin chain model
suitable to describe  azurite at high magnetic fields.
Interesting features of these highly frustrated spin models consist
in a steep increase of the entropy at very small temperatures and a characteristic extra low-temperature peak in the specific heat.
The most prominent effect is the existence of a magnetic-field driven Berezinskii-Kosterlitz-Thouless phase transition 
occurring in the two-dimensional model.
\end{abstract}

\maketitle

\section{Introduction}
\label{sec1}
\setcounter{equation}{0}

The study of frustrated quantum antiferromagnets is one of the most active research fields in condensed matter physics.\cite{lnp}
Among them there is a wide class of one-, two-, and three-dimensional frustrated quantum Heisenberg antiferromagnets 
with dispersionless (flat) lowest magnon bands.
There is currently a great deal of general interest in flat-band systems, 
since new many-body phases can be realized there
(see Refs.~\onlinecite{localized_magnons1,optical1,optical2,alter_1,alter_2,lm,SP,topological,topological_tasaki,prl2012} and references therein).

Interestingly flat-band {\it quantum} spin systems\cite{lm,SP} admit the application of specific methods of {\it classical} statistical mechanics 
to  study their high-field low-temperature behavior. 
For the application of classical statistical mechanics on the quantum systems 
the concept of many-body independent localized-magnon states is crucial.\cite{lm,localized_magnons1,SP,localized_magnons2,epjb2006,bilayer,double_tetrahedra_epjb}
Typical ground-state features related to the localized-magnon states are 
zero-temperature magnetization plateaus and jumps,\cite{lm}
high-field spin-Peierls lattice instabilities,\cite{SP}
and a residual ground-state entropy at the saturation field.\cite{localized_magnons1,zhito_hon2004,localized_magnons2,epjb2006}
Furthermore, 
these states set an additional low-energy scale that dominates the low-temperature thermodynamics in the vicinity of the saturation field resulting, 
e.g., in an extra peak in the specific heat at low temperatures.\cite{localized_magnons1,epjb2006}
In two-dimensional systems localized-magnon states may lead to a finite-temperature order-disorder phase transition 
of purely geometrical origin.\cite{localized_magnons1,bilayer}
It is worth mentioning that this concept for quantum spin systems 
is related to Mielke's and Tasaki's flat-band ferromagnetism of the Hubbard model.\cite{mielke,localized_electrons,prl2012}

The previously developed theories for localized-magnon spin lattices are valid 
if the conditions for localization of the magnon states are strictly fulfilled 
(so-called {\it ideal geometry}).
In real-life systems we cannot expect that, rather the violation of the localization condition is typical.
Therefore several questions arise:
What happens when the localization conditions are (slightly) violated?
Which features of the localized-magnon scenario survive? 
Which new effects may appear?

Although a systematic quantitative theory for such a case has not been elaborated until now,
it is in order to mention here 
Ref.~\onlinecite{distorted_ladder} considering a distorted frustrated two-leg spin ladder
(see also Ref.~\onlinecite{frbi3} related to a frustrated bilayer)
and 
Refs.~\onlinecite{effective_xy1} and \onlinecite{effective_xy2} dealing with a distorted diamond spin chain.
These studies, however, are not based on the localized-magnon picture and use a strong-coupling approach, see below.
In our recent paper\cite{drk} we touch this problem 
suggesting an heuristic ansatz for the partition function of a distorted diamond spin chain inspired by localized-magnon calculations.

The aim of the present paper is to develop a systematic treatment of a certain class of localized-magnon systems, 
namely the monomer class,\cite{epjb2006}
in the presence of small deviations from ideal  geometry. 
We consider three different frustrated spin lattices belonging to the monomer class, 
the diamond chain, the dimer-plaquette chain (Fig.~\ref{fig01}), and the square-kagome lattice (Fig.~\ref{fig02}).
We mention that these frustrated quantum antiferromagnets were discussed previously in the literature by various authors.\cite{diamond,dim_pla,sqkag}

Our strategy is to eliminate high-energy degrees of freedom, 
this way constructing low-energy effective Hamiltonians which are much simpler to treat than the initial ones.
Among the considered lattices the two-dimensional square-kagome lattice is particularly interesting, 
since it shares some properties with the kagome lattice.\cite{sqkag}
Moreover, generally in two dimensions a richer phase diagram can be expected.
The square-kagome lattice also admits a straightforward application of the strong-coupling method developed in previous papers, 
see, e.g., Refs.~\onlinecite{effective_xy1,effective_xy2,brenig}.
Note, however, that the  strong-coupling approach is not custom-tailored to the problem at hand, 
since it does not take advantage of the special localized-magnon properties.  
We also want to compare our theoretical findings with available experimental results for azurite.\cite{Kikuchi}
Azurite is known to be the model compound for a diamond spin chain
(for other compounds with diamond structure, see Ref.~\onlinecite{diamondchains_ssrealizations}).
Although its exchange parameter set  differs from the ideal localized-magnon geometry,
it is not too far from it.\cite{azurite-parameters,effective_xy2}

The rest of the paper is organized as follows.
First we introduce the models and briefly illustrate the localized-magnon
scenario (Sec.~\ref{sec2}).
Then we construct effective Hamiltonians (Sec.~\ref{sec3})
considering separately the strong-coupling approach and the localized-magnon based approach.
In Sec.~\ref{sec3} we compare exact-diagonalization results for the full and the
corresponding effective models 
to estimate the validity of the effective models.
In Sec.~\ref{sec4} we use the constructed low-energy effective models 
to discuss the low-temperature properties of the initial frustrated quantum antiferromagnets at high fields.
We summarize our findings in Sec.~\ref{sec5}.

\section{Models. Independent localized magnons}
\label{sec2}
\setcounter{equation}{0}

\begin{figure}
\begin{center}
\includegraphics[clip=on,width=70mm,angle=0]{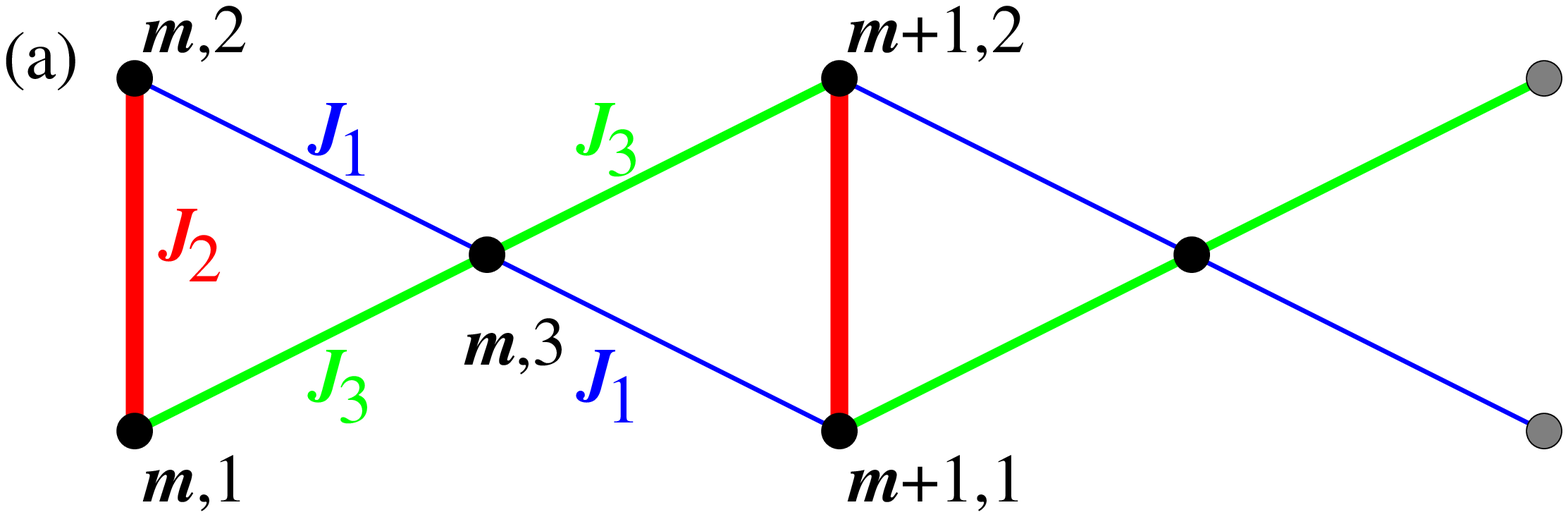}\\
\vspace{5mm}
\includegraphics[clip=on,width=82mm,angle=0]{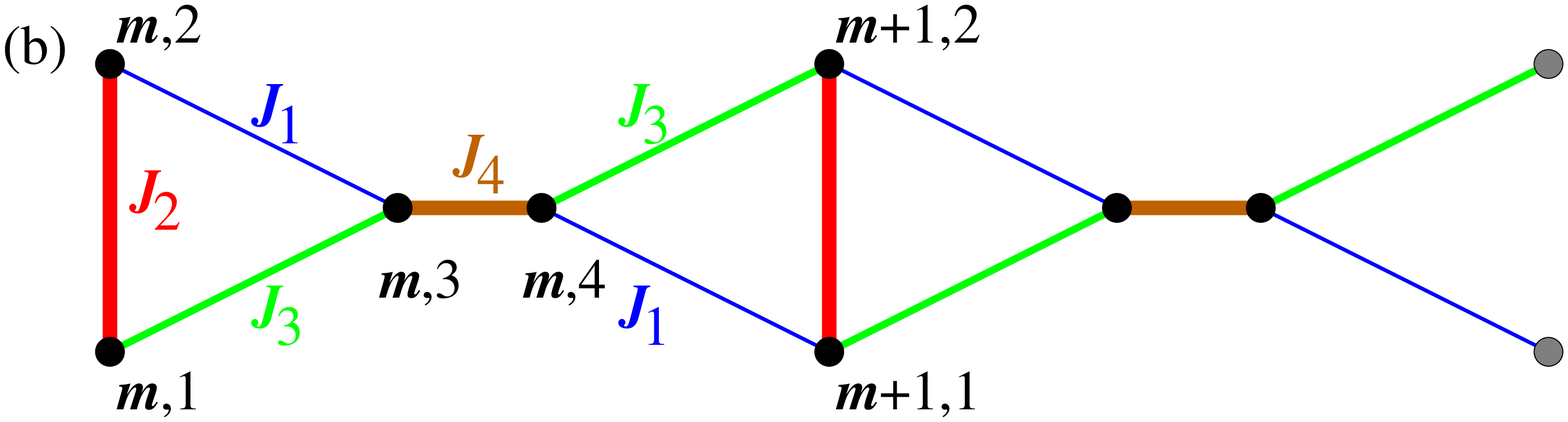}
\caption
{(Color online) 
(a) The diamond chain and (b) the dimer-plaquette chain described by Hamiltonian (\ref{201}).
The trapping cells (vertical dimers) for localized magnons are indicated by bold red lines ($J_2$ bonds).}
\label{fig01}
\end{center}
\end{figure}

\begin{figure}
\begin{center}
\includegraphics[clip=on,width=83mm,angle=0]{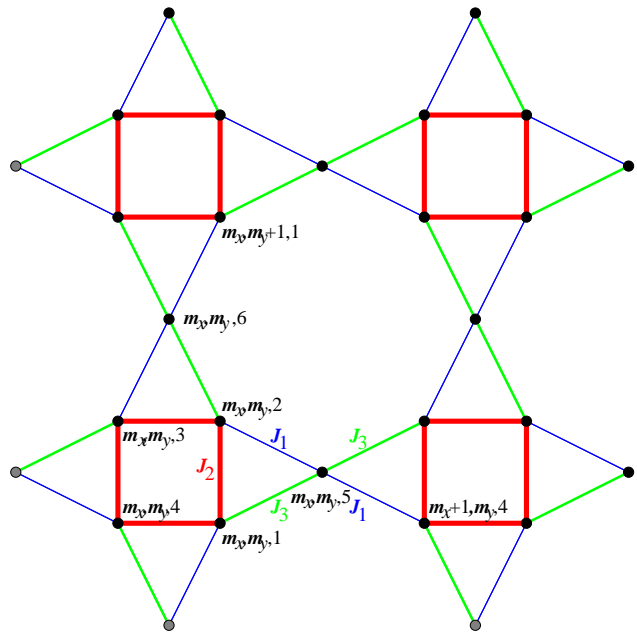}
\caption
{(Color online) 
The square-kagome lattice described by Hamiltonian (\ref{201}).
The trapping cells (squares) for localized magnons are indicated by bold red lines ($J_2$ bonds).}
\label{fig02}
\end{center}
\end{figure}

In the present study
we consider the standard spin-1/2 antiferromagnetic Heisenberg model in a magnetic field 
with the Hamiltonian
\begin{eqnarray}
\label{201}
H=\sum_{(ij)} J_{ij} {\bf{s}}_i \cdot {\bf{s}}_j-hS^z,
\;\;
S^z=\sum_{i=1}^Ns_i^z, \;\; J_{ij} > 0.
\end{eqnarray}
Here the first sum runs over all nearest-neighbor bonds on a lattice, whereas the second one runs over all $N$ lattice sites.
Note that $[S^z,H]=0$, i.e., the eigenvalues of $S^z$ are good quantum numbers.
The pattern of the exchange integrals $J_{ij}$ of the three different frustrated lattices considered here 
is shown in detail  in Figs.~\ref{fig01} and \ref{fig02}.
On a particular  lattice
a localized-magnon state can be located within a characteristic trapping cell
due to destructive quantum interference.\cite{lm}
For the diamond and dimer-plaquette chains the trapping cells are the vertical dimers, see Fig.~\ref{fig01}, 
for the square-kagome lattice the trapping cell is a square, see Fig.~\ref{fig02}. 
Owing to the localized nature of these states 
the many-magnon states in the subspaces $S^z=N/2-2,\ldots N/2-n_{\max}$, $n_{\max}\propto N$ 
can be constructed by filling the traps by localized magnons. 
Clearly, all these states are linear independent.\cite{linear_in}
Moreover, it can be shown that these localized-magnon states have the lowest energy in their  corresponding $S^z$-subspace,  
if the strength of the antiferromagnetic bonds of the trapping cells $J_2$ exceeds a lower bound.\cite{lm,Schmidt}

The degeneracy of the localized-magnon states are calculated 
via mapping onto spatial configuration of corresponding hard-core objects on an auxiliary lattice.\cite{localized_magnons1,localized_magnons2,epjb2006}
For the frustrated quantum antiferromagnet on the lattices at hand 
this hard-core system is a classical gas of monomers on a chain or a square lattice.\cite{localized_magnons2,epjb2006}

It is important to notice that the magnon localization occurs due to the specific lattice geometry
and hence requires certain relations between the bonds $J_{ij}$.
For the considered traps (single bond or square) this condition is fulfilled 
if an arbitrary bond of the trapping cell and the surrounding bonds attached to the two sites of this bond form an isosceles triangle,
i.e., $J_1=J_3=J$ in Figs.~\ref{fig01} and \ref{fig02}.
At low temperatures and for magnetic fields $h$ around the saturation field $h_{\rm{sat}}=h_1$
($h_1=J_2+J$ for the diamond and dimer-plaquette chains and  $h_1=2J_2+J$ for the square-kagome lattice)
the contribution of localized states is dominating in the partition function.\cite{epjb2006}
In the present study we deal with the case when the localization conditions are violated.
To be specific, 
in what follows we consider a violation of the ideal geometry 
by taking into account different values of $J_1$ and $J_3$ but fixing their average, 
i.e., $J_1\ne J_3$, $J_1+J_3=2J$, 
see Figs.~\ref{fig01} and \ref{fig02}.
This choice is relevant for azurite\cite{azurite-parameters,effective_xy2} 
and it is appropriate to illustrate the main point of our considerations.

For the further treatment of the models we introduce a convenient labeling of the lattice sites by a pair of indeces, 
where the first number enumerates the cells
[$m=1,\ldots,{\cal{N}}$, 
${\cal{N}} = N/3$ for the diamond chain,
${\cal{N}} = N/4$ for the dimer-plaquette chain,
${\cal{N}} = N/6$ for the square-kagome lattice,
$N$ is the number of sites;
for the square-kagome lattice the cells are enumerated by the vector index ${\bf{m}}=(m_x,m_y)$]
and the second one enumerates the position of the site within the cell, 
see Figs.~\ref{fig01} and \ref{fig02}.

\section{Effective Hamiltonians}
\label{sec3}
\setcounter{equation}{0}

\subsection{Strong-coupling approach}
\label{strong_coupling}

The strong-coupling perturbation theory is well established in the theory of quantum spin systems, 
see, e.g., Refs.~\onlinecite{effective_xy1,effective_xy2,brenig}.
Note that it is not necessarily related to the existence of localized-magnon states.
We begin with a brief illustration of the main steps of such an analysis.
The strong-coupling approach starts from finite elements/cells 
(e.g., dimers or squares) 
of the spin lattice which do not have common sites and have sufficiently large couplings, here $J_2$. 
The number of cells is denoted by ${\cal N}$,  ${\cal N}/N<1$.
On these cells the spin problem is solved analytically. 
(In the context of localized-magnon states these cells will play the role of the trapping cells.)
The cells are joined via weaker connecting bonds $J_i$, $i\ne 2$.
The strong-coupling approach is based on the assumption that the coupling $J_2$ is the dominant one, 
i.e., $J_i/J_2\ll 1$, $i\ne 2$.
Specifically, 
at high fields considered here only a few states of the trapping cell are relevant,
namely,
the fully polarized state $\vert u\rangle$ and the one-magnon state $\vert d\rangle$.
All other sites have fully polarized spins.
As the magnetic field decreases from very large values, 
the ground state of the cell undergoes a transition between the state $\vert u\rangle$ and the state $\vert d\rangle$ 
at the ``bare'' saturation field $h_0$ of a cell.
The Hamiltonian $H$ is splitted into 
a ``main'' part $H_{\rm{main}}$
[the Hamiltonian of all cells and the Zeeman interaction of all spins with the magnetic field $h_0$;
for the dimer-plaquette chain it includes also the interaction along horizontal
bonds, $J_4$, see Fig.~\ref{fig01}(b)]
and 
a perturbation $V$
(the rest of the Hamiltonian $H$).
The ground state $\vert\varphi_0\rangle$ of the Hamiltonian without the connecting bonds, 
i.e., $J_i=0$ for $i\ne 2$, 
at $h-h_0=0$ is $2^{{\cal{N}}}$-fold degenerate 
and forms a model space defined by the projector $P=\vert\varphi_0\rangle\langle\varphi_0\vert$.
For $J_i \ne 0$, $i\ne 2$ and $h-h_0 \ne 0$ we construct an effective Hamiltonian $H_{\rm{eff}}$ 
which acts on the model space only but gives the exact ground-state energy. 
$H_{\rm{eff}}$ can be found perturbatively and is given by\cite{klein,fulde,essler}
\begin{eqnarray}
\label{301}
H_{\rm{eff}}
=PHP+PV\sum_{\alpha\ne 0}\frac{\vert \varphi_{\alpha}\rangle\langle \varphi_{\alpha}\vert}{\varepsilon_0-\varepsilon_{\alpha}}VP+\ldots .
\end{eqnarray}
Here $\vert \varphi_{\alpha}\rangle$, $\alpha\ne 0$, are excited states of $H_{\rm{main}}$.
Finally, to rewrite the effective Hamiltonian in a more transparent form amenable for further analysis
it might be convenient to introduce (pseudo)spin operators representing the states of each trapping cell.

Next we present the concrete effective Hamiltonians obtained according to the above described procedure 
for the frustrated quantum Heisenberg antiferromagnets at hand.

\subsubsection{Diamond chain}

In the case of the diamond chain
the following two states of each vertical bond are taken into account:
$\vert u\rangle=\vert\uparrow_1\uparrow_2\rangle$ 
with the energy $J_2/4-h$
and 
$\vert d\rangle=\left(\vert \uparrow_1\downarrow_2\rangle-\vert \downarrow_1\uparrow_2\rangle\right)/\sqrt{2}$ 
with the energy $-3J_2/4$. 
Furthermore, $h_0=J_2$.
For the projector onto the ground-state manifold we have
\begin{eqnarray}
\label{302}
P=\otimes_{m=1}^{{\cal{N}}}P_m,
\nonumber\\
P_m={\cal{P}}_{m}\otimes \left(\vert\uparrow_3\rangle\langle\uparrow_3\vert\right)_m,
{\cal{P}}_{m}=\left(\vert u\rangle\langle u\vert+\vert d\rangle\langle d\vert\right)_{m}.
\;\;
\end{eqnarray}
The set of relevant excited states $\vert\varphi_{\alpha}\rangle$, $\alpha\ne 0$,
is  the set of the ${\cal{N}} 2^{\cal{N}}$ states 
with only one flipped spin $\vert\downarrow_3\rangle$ on those sites which connect two neighboring vertical bonds.
Moreover, $\varepsilon_\alpha=\varepsilon_0+h_0=\varepsilon_0+J_2$.
After introducing (pseudo)spin-1/2 operators for each vertical bond 
\begin{eqnarray}
\label{303}
T^z=\frac{1}{2}\left( \vert u\rangle\langle u\vert - \vert d\rangle\langle d\vert\right),
T^+=\vert u\rangle\langle d\vert,
T^-=\vert d\rangle\langle u\vert
\end{eqnarray}
Eq.~(\ref{301}) becomes
\begin{eqnarray}
\label{304}
H_{{\rm{eff}}}
=
\sum_{m=1}^{{\cal{N}}}
\left[ {\sf{J}}\left(T_m^xT_{m+1}^x + T_m^yT_{m+1}^y\right) -{\sf{h}}T^z_m + {\sf{C}} \right]
\end{eqnarray}
with
\begin{eqnarray}
\label{305}
{\sf{J}}=\frac{\left(J_1-J_3\right)^2}{4J_2},
\nonumber\\
{\sf{h}}=h-h_1-\frac{\left(J_1-J_3\right)^2}{4J_2},
\nonumber\\
{\sf{C}}=-h-\frac{J_2}{4}+\frac{J}{2}-\frac{\left(J_1-J_3\right)^2}{8J_2},
\nonumber\\
h_1=J_2+J,
\,\,\,
J=\frac{J_1+J_3}{2}.
\end{eqnarray}
The effective Hamiltonian in strong-coupling approach corresponds to an unfrustrated spin-1/2 isotropic $XY$ chain in a transverse magnetic field.\cite{lieb}
This result coincides with that one obtained in Ref.~\onlinecite{effective_xy2}.

\subsubsection{Dimer-plaquette chain}

In the case of the dimer-plaquette chain the relevant two states of each vertical bond are the same,
but the projector onto the ground-state manifold obviously contains the projector 
on the up-state for both spins on each horizontal bond 
[$J_4$ bond in Fig.~\ref{fig01}(b)],
i.e., 
$P_m={\cal{P}}_{m}\otimes\left(\vert\uparrow_3\rangle\langle\uparrow_3\vert\otimes\vert\uparrow_4\rangle\langle\uparrow_4\vert\right)_m$,
where ${\cal{P}}_{m}$ is defined in Eq.~(\ref{302}). 
Instead of a flipped spin on the site which connects two neighboring vertical bonds being an excited state for the diamond chain,
we consider here two classes of excited states.
The first one contains the singlet state on the horizontal bond, 
$\vert s\rangle=\left(\vert\uparrow_3\downarrow_4\rangle-\vert\downarrow_3\uparrow_4\rangle\right)/\sqrt{2}$.
The energy of this excited state is $\varepsilon_\alpha=\varepsilon_0+J_2-J_4$.
The second one contains the $S^z=0$ component of the triplet state on the horizontal bond, 
$\vert t\rangle=\left(\vert\uparrow_3\downarrow_4\rangle+\vert\downarrow_3\uparrow_4\rangle\right)/\sqrt{2}$.
The energy of this excited state is $\varepsilon_\alpha=\varepsilon_0+J_2$.
As a result,
we again obtain the one-dimensional spin-1/2 isotropic $XY$ model in a transverse field given by  Eq.~(\ref{304}), however, with different parameters,
\begin{eqnarray}
\label{306}
{\sf{J}}=-\frac{\left(J_1-J_3\right)^2}{8\left(J_2-J_4\right)}\frac{J_4}{J_2},
\nonumber\\
{\sf{h}}=h-h_1-\frac{\left(J_1-J_3\right)^2}{8\left(J_2-J_4\right)}\frac{2J_2-J_4}{J_2},
\nonumber\\
{\sf{C}}=-\frac{3}{2}h-\frac{J_2}{4}+\frac{J_4}{4}+\frac{J}{2}
-\frac{\left(J_1-J_3\right)^2}{16\left(J_2-J_4\right)}\frac{2J_2-J_4}{J_2},
\nonumber\\
h_1=J_2+J,
\,\,\,
J=\frac{J_1+J_3}{2}. \;\;
\end{eqnarray}
Note that we get ${\sf{J}}\to 0$ if $J_4\to 0$ as it is expected from physical arguments.

\subsubsection{Square-kagome lattice}

In the case of the square-kagome lattice
we take into account the following two states of each square:
$\vert u\rangle=\vert \uparrow_1\uparrow_2\uparrow_3\uparrow_4\rangle$ with the energy $J_2-2h$
and 
$\vert d\rangle=\left(\vert \uparrow_1\uparrow_2\uparrow_3\downarrow_4\rangle 
-\vert \uparrow_1\uparrow_2\downarrow_3\uparrow_4\rangle 
+\vert \uparrow_1\downarrow_2\uparrow_3\uparrow_4\rangle 
-\vert \downarrow_1\uparrow_2\uparrow_3\uparrow_4\rangle\right)/2$ with the energy $-J_2-h$,
and $h_0=2J_2$.
For the projector $P_m$ in Eq.~(\ref{302}) we now have
$P_{\bf{m}}={\cal{P}}_{\bf{m}}\otimes \left(\vert\uparrow_5\rangle\langle\uparrow_5\vert \otimes \vert\uparrow_6\rangle\langle\uparrow_6\vert\right)_{\bf{m}}$,
${\cal{P}}_{\bf{m}}=\left(\vert u\rangle\langle u\vert+\vert d\rangle\langle d\vert\right)_{\bf{m}}$,
see also Fig.~\ref{fig02}.
Similar to the case of the diamond chain 
the relevant excitations have one flipped spin on those sites connecting two neighboring squares either in the horizontal or in the vertical direction.
Their energy is $\varepsilon_\alpha=\varepsilon_0+h_0=\varepsilon_0+2J_2$.
The resulting effective Hamiltonian (\ref{301}) has the form
\begin{eqnarray}
\label{307}
H_{{\rm{eff}}}={\sf{J}}\sum_{(mn)}\left(T_m^xT_{n}^x + T_m^yT_{n}^y\right)-{\sf{h}}\sum_{m=1}^{{\cal{N}}}T^z_m
+{\cal{N}}{\sf{C}},
\end{eqnarray}
where the first sum runs over all nearest neighbors on a square lattice
and
\begin{eqnarray}
\label{308}
{\sf{J}}=-\frac{\left(J_1-J_3\right)^2}{16J_2},
\nonumber\\
{\sf{h}}=h-h_1-\frac{\left(J_1-J_3\right)^2}{8J_2},
\nonumber\\
{\sf{C}}=-\frac{5}{2}h+\frac{3}{2}J-\frac{\left(J_1-J_3\right)^2}{16J_2},
\nonumber\\
h_1=2J_2+J,
\,\,\,
J=\frac{J_1+J_3}{2}.
\end{eqnarray}
Again we get a basic model of quantum magnetism, 
the square-lattice spin-1/2 isotropic $XY$ model in a transverse magnetic field. 

\subsection{Localized-magnon approach}
\label{localized_magnon}

As illustrated above, 
the strong-coupling consideration provides a low-energy description for the considered frustrated quantum Heisenberg antiferromagnets at high fields
provided that the intracell coupling $J_2$ is much larger than all other couplings.
This might be a quite natural requirement for cell-based (e.g., dimer-based or square-based) spin systems
but it is not necessary in the context of localized-magnon systems.
Indeed, the localized-magnon picture for the considered spin systems emerges 
when $J_1-J_3=0$ whereas $\left(J_1+J_3\right)/2=J$ is only smaller, but not much smaller than $J_2$.\cite{epjb2006}

At first glance, 
we may extend straightforwardly the above described scheme introducing another splitting of the Hamiltonian $H$
taking advantage of the specific features of the localized-magnon scenario.
As the main part of the Hamiltonian we take that part of the initial Hamiltonian that corresponds to the ideal geometry, 
i.e., a Hamiltonian with $J=\left(J_1+J_3\right)/2$ instead of $J_1$ and $J_3$ at $h=h_1$.
The remaining part of the initial Hamiltonian, which contains $J_i-J$ and $h-h_1$ only, we consider as the perturbation.
The ground state of the main part is the well known set of independent localized-magnon states,
however, the excited states which are necessary for calculation of the second term in Eq.~(\ref{301}) are generally unknown.

We may overcome this difficulty
considering as a starting point instead of $H$ the Hamiltonian
\begin{eqnarray}
\label{309}
{\cal{H}}={\cal{P}}H{\cal{P}},
\;\;\;
{\cal{P}}=\otimes_{m=1}^{{\cal{N}}}{\cal{P}}_m,
\end{eqnarray}
where ${\cal{P}}_m$ is the projector on the relevant states of the trapping cell $m$.
It was defined for the various models under consideration in Sec.~\ref{strong_coupling}.
It is important to note that by contrast  to the projection operator $P$ used in strong-coupling approximation [Eqs.~(\ref{301}) and (\ref{302})],
the  projection operator $\cal{P}$ used here does not fix the intermediate spins connecting the trapping cells thus allowing more degrees of freedom.
Nevertheless, we have made an approximation by reducing the number of states taken into account 
[namely, 
instead of 4 (16) states of each vertical bond (square) we consider now only 2 of them,
the fully polarized state $\vert u\rangle$ and the localized-magnon state $\vert d\rangle$].
Moreover, the restriction to  states $\vert u\rangle$ and $\vert d\rangle$ limits the possibility of spreading the cell states over the lattice.
For the reduced set of degrees of freedom it is then straightforward to introduce in ${\cal{H}}$ again (pseudo)spin operators.

The usefulness of this new Hamiltonian ${\cal{H}}$ is twofold.
First, it is interesting in its own rights 
providing an effective description of the initial spin model in terms of a simpler model with a smaller number of sites, 
see below.
Second, a major advantage is that for ${\cal{H}}$ we can eliminate the spin variables relating to the sites which connect the traps 
perturbatively with respect to small deviations from the ideal geometry 
(but not with respect to the total strength of the connecting bonds)
arriving at an effective model which is certainly an improvement of the strong-coupling one.
More specifically,
we split ${\cal{H}}$ into a main part ${\cal{H}}_{\rm{main}}$ 
(that is the Hamiltonian ${\cal{H}}$ with $J_1=J_3=J$ and $h=h_1$)
and a perturbation ${\cal{V}}={\cal{H}}-{\cal{H}}_{\rm{main}}$.
The ground state of ${\cal{H}}_{\rm{main}}$ is the same as that of the strong-coupling approach $\vert\varphi_0\rangle$
(although with another value of the ground-state energy $\varepsilon_0$,
since the connecting bonds are present in ${\cal{H}}_{\rm{main}}$)
and, hence, also the projector is the same $P=\vert\varphi_0\rangle\langle\varphi_0\vert$.
Moreover, 
for the Hamiltonian ${\cal{H}}_{\rm{main}}$ all relevant excited states $\vert\varphi_\alpha\rangle$, $\alpha\ne 0$, are known 
and therefore an effective Hamiltonian (\ref{301}) can be worked out.

Below we present further details for each frustrated quantum Heisenberg antiferromagnet under consideration.

\subsubsection{Diamond chain}

After eliminating irrelevant states of the vertical bond in favor of the two relevant ones, 
$\vert u\rangle$ and $\vert d\rangle$, 
and introducing pseudospin $T$-operators (\ref{303})
we obtain from Eq.~(\ref{309})
\begin{eqnarray}
\label{310}
{\cal{H}}
=\sum_{m=1}^{{\cal{N}}}
\left[ 
-\frac{h}{2}-\frac{J_2}{4}-\left(h-J_2\right)T^z_m -\left(h-J\right)s^z_{m,3} 
\right.
\nonumber\\
\left.
+\frac{J_1-J_3}{\sqrt{2}} \left(T^x_ms^x_{m,3}+T^y_ms^y_{m,3}\right)
+JT^z_ms^z_{m,3}
\right.
\nonumber\\
\left.
-\frac{J_1-J_3}{\sqrt{2}} \left(s^x_{m,3}T^x_{m+1}+s^y_{m,3}T^y_{m+1}\right)
+Js^z_{m,3}T^z_{m+1}
\right].
\;\;\;\;\;
\end{eqnarray}
That is a spin-1/2 $XXZ$ model with alternating isotropic $XY$ bonds in an alternating magnetic $z$-field 
on a simple chain of $2{\cal{N}}$ sites, 
i.e., the unit cell contains two sites.

To exclude further the spin variables at the sites $m,3$, $m=1,\ldots,{\cal{N}}$ perturbatively,
we consider the main Hamiltonian
\begin{eqnarray}
\label{311}
{\cal{H}}_{\rm{main}}
=\sum_{m=1}^{{\cal{N}}}
\left[ 
-\frac{h_1}{2}-\frac{J_2}{4}-\left(h_1-J_2\right)T^z_m -\left(h_1-J\right)s^z_{m,3} 
\right.
\nonumber\\
\left.
+J\left(T^z_ms^z_{m,3}+s^z_{m,3}T^z_{m+1}\right)
\right],
\;\;\;\;\;
\end{eqnarray}
where $h_1=J_2+J$.
That is an Ising chain Hamiltonian with known eigenstates $\vert\varphi_\alpha\rangle$.
The set of ground states $\prod_m\left(\vert{\rm{v}}\rangle\vert\uparrow_3\rangle\right)_m$,
where ${\rm{v}}$ is either $u$ or $d$, 
has the energy $\varepsilon_0=(-5J_2/4-J/2){\cal{N}}$.
Now we consider the perturbation ${\cal{V}}={\cal{H}}-{\cal{H}}_{\rm{main}}$ 
and the set of excited states $\vert\varphi_\alpha\rangle$ which enter Eq.~(\ref{301}).
Again we consider the states with one flipped spin on the site $m,3$.
However, now (since $J\ne 0$) the energy of the excited state depends on the states of the neighboring vertical bonds. 
Namely, 
$\varepsilon_\alpha=\varepsilon_0+J_2-J$ if for the neighboring sites $T^z=1/2$,
$\varepsilon_\alpha=\varepsilon_0+J_2$ if for one of the neighboring sites $T^z=1/2$ and for another one $T^z=-1/2$,
and
$\varepsilon_\alpha=\varepsilon_0+J_2+J$ if for the neighboring sites $T^z=-1/2$.
Taking into account this circumstance in Eq.~(\ref{301}),
after straightforward calculations we arrive at the effective Hamiltonian
\begin{eqnarray}
\label{312}
{\cal{H}}_{{\rm{eff}}}=\sum_{m=1}^{{\cal{N}}}
\left[{\sf{J}}\left(T_m^xT_{m+1}^x + T_m^yT_{m+1}^y\right) +{\sf{J}}^zT_m^zT_{m+1}^z
\right.
\nonumber\\
\left.
-{\sf{h}}T^z_m +{\sf{C}} \; \right ]\quad
\end{eqnarray}
with the following parameters
\begin{eqnarray}
\label{313}
{\sf{J}}=\frac{\left(J_1-J_3\right)^2}{4J_2}\frac{1}{1-\frac{J}{J_2}},
\nonumber\\
{\sf{J}}^z=\frac{\left(J_1-J_3\right)^2}{4J_2}
\left(\frac{1}{1-\frac{J}{J_2}}-1\right),
\nonumber\\
{\sf{h}}=h-h_1-\frac{\left(J_1-J_3\right)^2}{4J_2},
\nonumber\\
{\sf{C}}=-h-\frac{J_2}{4}+\frac{J}{2}
-\frac{\left(J_1-J_3\right)^2}{16J_2}
\left(\frac{1}{1-\frac{J}{J_2}}+1\right),
\nonumber\\
J=\frac{J_1+J_3}{2},
\,\,\,
h_1=J_2+J. \quad
\end{eqnarray}

We may expand the effective couplings and field with respect to $J/J_2$ in Eq.~(\ref{313}):
\begin{eqnarray}
\label{314}
{\sf{J}}=\frac{\left(J_1-J_3\right)^2}{4J_2}\left(1+\frac{J}{J_2}+\ldots\right),
\nonumber\\
{\sf{J}}^z=\frac{\left(J_1-J_3\right)^2}{4J_2}\left(\frac{J}{J_2}+\ldots\right),
\nonumber\\
{\sf{h}}=h-h_1-\frac{\left(J_1-J_3\right)^2}{4J_2}.
\end{eqnarray}
This result reproduces not only the second-order perturbation theory in $1/J_2$ reported in Refs.~\onlinecite{effective_xy1} and \onlinecite{effective_xy2},
but also the third-order terms \cite{effective_xy1} in $1/J_2$,
see Eq.~(6.3) of Ref.~\onlinecite{effective_xy1}.

\subsubsection{Dimer-plaquette chain}

In the case of the dimer-plaquette chain,
the Hamiltonian given in Eq.~(\ref{309}) reads:
\begin{eqnarray}
\label{315}
{\cal{H}}=\sum_{m=1}^{{\cal{N}}}
\left[ 
-\frac{h}{2}-\frac{J_2}{4}\qquad \qquad \qquad 
\right.
\nonumber\\
\left.
-\left(h-J_2\right)T^z_m -\left(h-\frac{J}{2}\right)\left(s^z_{m,3}+s^z_{m,4}\right) \qquad
\right.
\nonumber\\
\left.
+\frac{J_1-J_3}{\sqrt{2}}\left(T^x_ms^x_{m,3}+T^y_ms^y_{m,3}\right) +JT^z_ms^z_{m,3} \qquad 
\right.
\nonumber\\
\left.
+ J_4{\bf{s}}_{m,3}\cdot{\bf{s}}_{m,4}\qquad 
\right.
\nonumber\\
\left.
-\frac{J_1-J_3}{\sqrt{2}}\left(s^x_{m,4}T^x_{m+1}+s^y_{m,4}T^y_{m+1}\right)+Js^z_{m,4}T^z_{m+1}
\right]. 
\;\;\;\;\;
\end{eqnarray}
This Hamiltonian corresponds to a spin-1/2 $XXZ$ model in a magnetic $z$-field 
on a simple chain of $3{\cal{N}}$ sites
with a unit cell of three sites.

The starting point for the construction of the ${\cal{N}}$-site effective model is
\begin{eqnarray}
\label{316}
{\cal{H}}_{\rm{main}}=\sum_{m=1}^{{\cal{N}}}
\Big [ 
-\frac{h_1}{2}-\frac{J_2}{4}
\nonumber\\
-\left(h_1-J_2\right)T^z_m -\left(h_1-\frac{J}{2}\right)\left(s^z_{m,3}+s^z_{m,4}\right) 
\nonumber\\
+JT^z_ms^z_{m,3} + J_4{\bf{s}}_{m,3}\cdot{\bf{s}}_{m,4} + Js^z_{m,4}T^z_{m+1}
\Big ]
\end{eqnarray}
with $h_1=J_2+J$.
That is an Ising-Heisenberg chain Hamiltonian,\cite{ising-heisenberg}
cf. also Eq.~(\ref{311}).
Next, we take into account the perturbation ${\cal{V}}$ and consider excited states $\vert\varphi_\alpha\rangle$.
In contrast to the strong-coupling treatment 
where the excited states correspond to one horizontal bond in the state $\vert s\rangle$ or $\vert t\rangle$,
now we are faced with a more complicated situation, 
since the excited states and their energies depend on the states of the neighboring vertical bonds. 
If both neighboring vertical bonds are in the state with $T^z=1/2$, 
the first excited state 
$\vert\ldots\left(\vert u\rangle\vert s\rangle\right)_m 
\left(\vert u\rangle\vert\uparrow_3\uparrow_4\rangle\right)_{m+1}\ldots\rangle$
has the energy $\varepsilon_\alpha=\varepsilon_0+J_2-J_4$,
and the other excited state 
$\vert\ldots\left(\vert u\rangle\vert t\rangle\right)_m \left(\vert u\rangle\vert\uparrow_3\uparrow_4\rangle\right)_{m+1}\ldots\rangle$
has the energy $\varepsilon_\alpha=\varepsilon_0+J_2$.
Similarly,
if both neighboring vertical bonds have $T^z=-1/2$,
the excited state 
$\vert\ldots\left(\vert d\rangle\vert s\rangle\right)_m \left(\vert d\rangle\vert
\uparrow_3\uparrow_4\rangle\right)_{m+1}\ldots\rangle$
has the energy $\varepsilon_\alpha=\varepsilon_0+J_2-J_4+J$,
and the excited state 
$\vert\ldots\left(\vert d\rangle\vert t\rangle\right)_m \left(\vert d\rangle\vert
\uparrow_3\uparrow_4\rangle\right)_{m+1}\ldots\rangle$
has the energy $\varepsilon_\alpha=\varepsilon_0+J_2+J$.
Furthermore,
if $T^z_m=1/2$ and $T^z_{m+1}=-1/2$,
the excited state 
$\vert\ldots\left(\vert u\rangle\vert s_{ud}\rangle\right)_m \left(\vert d\rangle\vert
\uparrow_3\uparrow_4\rangle\right)_{m+1}\ldots\rangle$,
$\vert s_{ud}\rangle=\left(\vert s\rangle+x\vert t\rangle\right)/\sqrt{1+x^2}$, 
$x=(J_4-\sqrt{J_4^2+J^2})/J$ 
has the energy $\varepsilon_\alpha=\varepsilon_0+J_2+J/2-J_4/2-\sqrt{J_4^2+J^2}/2$,
and the other excited state 
$\vert\ldots\left(\vert u\rangle\vert t_{ud}\rangle\right)_m \left(\vert d\rangle\vert
\uparrow_3\uparrow_4\rangle\right)_{m+1}\ldots\rangle$,
$\vert t_{ud}\rangle=\left(-x\vert s\rangle+\vert t\rangle\right)/\sqrt{1+x^2}$
has the energy $\varepsilon_\alpha=\varepsilon_0+J_2+J/2-J_4/2+\sqrt{J_4^2+J^2}/2$.
If $T^z_m=-1/2$ and $T^z_{m+1}=1/2$,
the excited state 
$\vert\ldots\left(\vert d\rangle\vert s_{du}\rangle\right)_m \left(\vert u\rangle\vert
\uparrow_3\uparrow_4\rangle\right)_{m+1}\ldots\rangle$,
$\vert s_{du}\rangle=\left(\vert s\rangle-x\vert t\rangle\right)/\sqrt{1+x^2}$
has the energy $\varepsilon_\alpha=\varepsilon_0+J_2+J/2-J_4/2-\sqrt{J_4^2+J^2}/2$,
whereas the excited state 
$\vert\ldots\left(\vert d\rangle\vert t_{du}\rangle\right)_m \left(\vert u\rangle\vert
\uparrow_3\uparrow_4\rangle\right)_{m+1}\ldots\rangle$,
$\vert t_{du}\rangle=\left(x\vert s\rangle+\vert t\rangle\right)/\sqrt{1+x^2}$
has the energy $\varepsilon_\alpha=\varepsilon_0+J_2+J/2-J_4/2+\sqrt{J_4^2+J^2}/2$.
Taking into account all these formulae in Eq.~(\ref{301}),
after some calculations we arrive again at the effective Hamiltonian given in Eq.~(\ref{312}),
however, now with the following parameters:
\begin{eqnarray}
\label{317}
{\sf{J}}= -\frac{\left(J_1-J_3\right)^2}{8\left(J_2-J_4\right)} +\frac{\left(J_1-J_3\right)^2}{8J_2},
\nonumber\\
{\sf{J}}^z=\frac{\left(J_1-J_3\right)^2}{8\left(J_2-J_4\right)}
\left[ 1 - \frac{1}{1+\frac{J+J_4-\sqrt{J_4^2+J^2}}{2\left(J_2-J_4\right)}} \frac{\left(1-x\right)^2}{1+x^2}\right]
\nonumber\\
+\frac{\left(J_1-J_3\right)^2}{8J_2}
\left[ 1 - \frac{1}{1+\frac{J-J_4+\sqrt{J_4^2+J^2}}{2J_2}} \frac{\left(1+x\right)^2}{1+x^2}\right],
\nonumber\\
{\sf{h}}=h-h_1
-\frac{\left(J_1-J_3\right)^2}{8\left(J_2-J_4\right)} 
\frac{1}{1+\frac{J+J_4-\sqrt{J_4^2+J^2}}{2\left(J_2-J_4\right)}}\frac{\left(1-x\right)^2}{1+x^2}
\nonumber\\
-\frac{\left(J_1-J_3\right)^2}{8J_2} 
\frac{1}{1+\frac{J-J_4+\sqrt{J_4^2+J^2}}{2J_2}}\frac{\left(1+x\right)^2}{1+x^2},
\nonumber\\
{\sf{C}}=-\frac{3}{2}h-\frac{J_2}{4}+\frac{J}{2}+\frac{J_4}{4}
\nonumber\\
-\frac{\left(J_1-J_3\right)^2}{32\left(J_2-J_4\right)}
\left[1 + \frac{1}{1+\frac{J+J_4-\sqrt{J_4^2+J^2}}{2\left(J_2-J_4\right)}}\frac{\left(1-x\right)^2}{1+x^2}\right]
\nonumber\\
-\frac{\left(J_1-J_3\right)^2}{32J_2}
\left[1 + \frac{1}{1+\frac{J-J_4+\sqrt{J_4^2+J^2}}{2J_2}}\frac{\left(1+x\right)^2}{1+x^2}\right],
\nonumber\\
J=\frac{J_1+J_3}{2},
\,\,\,
x=\frac{J_4-\sqrt{J_4^2+J^2}}{J},
\,\,\,
h_1=J_2+J. 
\quad
\end{eqnarray}
In the limiting case $J/J_2\to 0$, $J/J_4\to 0$ this result transforms into the one obtained using the strong-coupling approach,
cf. Eq.~(\ref{306}).

\subsubsection{Square-kagome lattice}

For the square-kagome lattice the Hamiltonian $\cal{H}$ (\ref{309}) reads 
\begin{widetext}
\begin{eqnarray}
\label{318}
{\cal{H}}=\sum_{{\bf{m}}}
\left[-\frac{3}{2}h
-\left(h-2J_2\right)T^z_{\bf{m}}-\left(h-\frac{3}{2}J\right)\left(s_{{\bf{m}},5}^z+s_{{\bf{m}},6}^z\right)
\right.
\nonumber\\
\left.
+\frac{J}{2}T^z_{\bf{m}}s_{{\bf{m}},5}^z+\frac{J_1-J_3}{2}\left(T^x_{\bf{m}}s_{{\bf{m}},5}^x+T^y_{\bf{m}}s_{{\bf{m}},5}^y\right)
+\frac{J}{2}s_{{\bf{m}},5}^zT^z_{m_x+1,m_y}+\frac{J_1-J_3}{2}\left(s_{{\bf{m}},5}^xT^x_{m_x+1,m_y}+s_{{\bf{m}},5}^yT^y_{m_x+1,m_y}\right)
\right.
\nonumber\\
\left.
+\frac{J}{2}T^z_{\bf{m}}s_{{\bf{m}},6}^z-\frac{J_1-J_3}{2}\left(T^x_{\bf{m}}s_{{\bf{m}},6}^x+T^y_{\bf{m}}s_{{\bf{m}},6}^y\right)
+\frac{J}{2}s_{{\bf{m}},6}^zT^z_{m_x,m_y+1}-\frac{J_1-J_3}{2}\left(s_{{\bf{m}},6}^xT^x_{m_x,m_y+1}+s_{{\bf{m}},6}^yT^y_{m_x,m_y+1}\right)
\right].
\end{eqnarray}
\end{widetext}
It corresponds to a spin-1/2 $XXZ$ model on a decorated square lattice 
(which is also known as the Lieb lattice\cite{lieb_lattice}),
see Fig.~\ref{fig02}.
The Hamiltonian ${\cal{H}}_{\rm{main}}$ is given by Eq.~(\ref{318}) with $J_1=J_3=J$ and $h=h_1=2J_2+J$.
The states with one flipped spin on those sites connecting two neighboring squares constitute the set of relevant excited states.
The energy of the excited states depends on the states of these two squares.
Namely, it acquires 
the value $\varepsilon_\alpha=\varepsilon_0+2J_2-J$ 
if both squares are in the $\vert u\rangle$ state,
the value $\varepsilon_\alpha=\varepsilon_0+2J_2-J/2$ 
if one of the squares is in the $\vert u\rangle$ state and the other one in the $\vert d\rangle$ state,
and 
the value $\varepsilon_\alpha=\varepsilon_0+2J_2$ 
if both squares are in the $\vert d\rangle$ state.
Taking this into account, 
we can calculate the second term of Eq.~(\ref{301}) and arrive at the Hamiltonian
\begin{eqnarray}
\label{319}
{\cal{H}}_{{\rm{eff}}}=\sum_{(mn)}
\left[{\sf{J}}\left(T_m^xT_{n}^x + T_m^yT_{n}^y\right) +{\sf{J}}^zT_m^zT_{n}^z\right]
\nonumber\\
-{\sf{h}}\sum_{m=1}^{{\cal{N}}}T^z_m +{\cal{N}}{\sf{C}}.
\end{eqnarray}
Here the first sum runs over the neighboring sites of an ${\cal{N}}$-site square lattice
and the parameters have the following values:
\begin{eqnarray}
\label{320}
{\sf{J}}=-\frac{\left(J_1-J_3\right)^2}{16J_2}\frac{1}{1-\frac{J}{2J_2}},
\nonumber\\
{\sf{J}}^z=\frac{\left(J_1-J_3\right)^2}{16J_2}
\left(\frac{1}{1-\frac{J}{2J_2}}-\frac{1}{1-\frac{J}{4J_2}}\right),
\nonumber\\
{\sf{h}}=h-h_1-\frac{\left(J_1-J_3\right)^2}{8J_2}\frac{1}{1-\frac{J}{4J_2}},
\nonumber\\
{\sf{C}}=-\frac{5}{2}h+\frac{3}{2}J
-\frac{\left(J_1-J_3\right)^2}{32J_2}
\left(\frac{1}{1-\frac{J}{2J_2}}+\frac{1}{1-\frac{J}{4J_2}}\right),
\nonumber\\
J=\frac{J_1+J_3}{2},
\,\,\,
h_1=2J_2+J.
\;\;\;\;\;
\end{eqnarray}
Obviously, in the limit $J/J_2 \to 0$ this result coincides with that one obtained within the strong-coupling approach,
cf. Eq.~(\ref{308}).

\subsection{Exact diagonalization}
\label{ED}

In this section we will discuss the region of validity of the effective Hamiltonians obtained in the previous sections 
by reducing the dimension of the full Hilbert space and by using perturbation expansions. 
To check the quality of the obtained effective Hamiltonians
we perform extensive exact diagonalization studies for the initial and the effective models
and compare the results focusing on magnetization curves and the temperature dependence of the specific heat.

\begin{figure}
\begin{center}
\includegraphics[clip=on,width=80mm,angle=0]{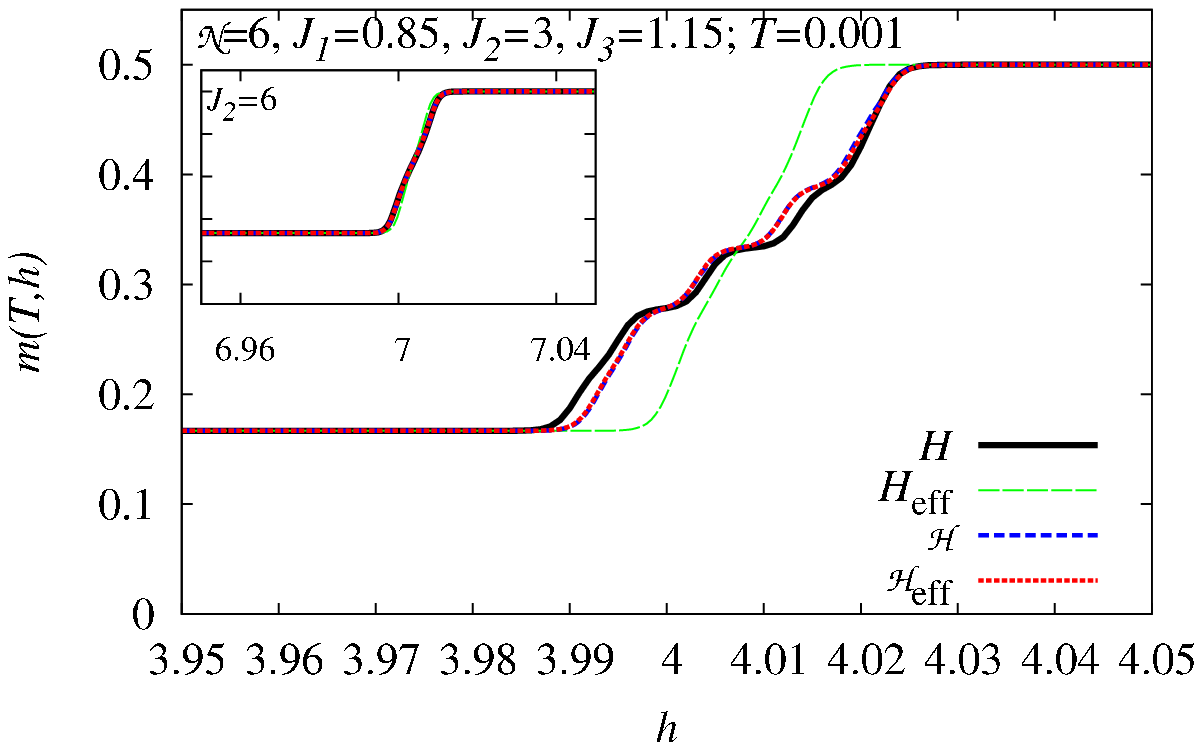}\\
\vspace{5mm}
\includegraphics[clip=on,width=80mm,angle=0]{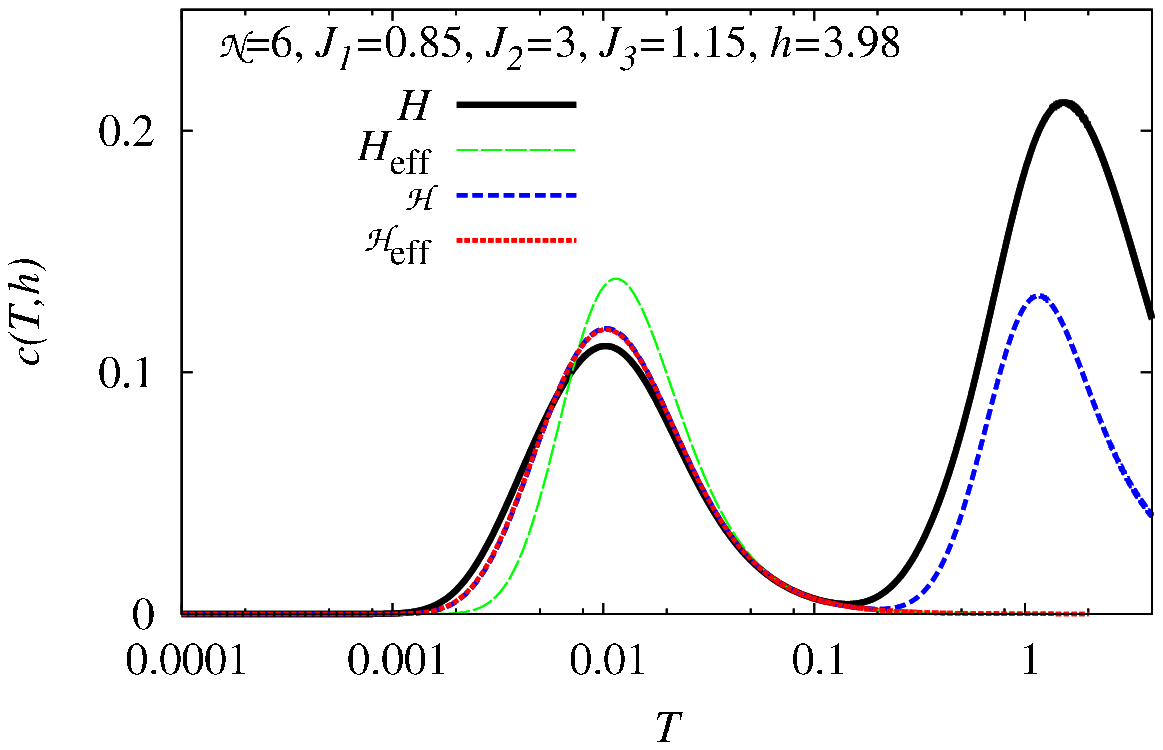}\\
\vspace{3mm}
\includegraphics[clip=on,width=80mm,angle=0]{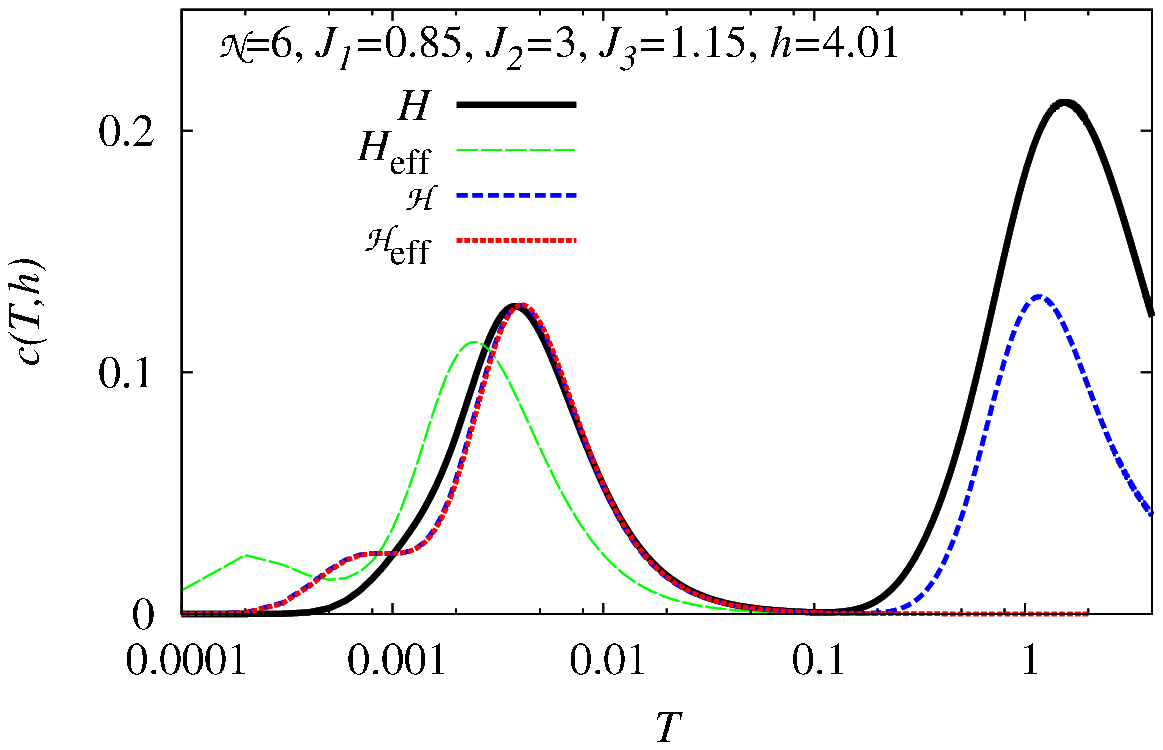}
\caption
{(Color online) 
Comparison of the full model with effective models for the distorted diamond chain of ${\cal{N}}=6$ cells: 
Field dependences of the magnetization (per site) and temperature dependences of the specific heat (per site) 
for $J_1=0.85$, $J_2=3$ (main panels) and $J_2=6$ (inset), $J_3=1.15$.
For more explanations, see the main text.}
\label{fig03}
\end{center}
\end{figure}

\begin{figure}
\begin{center}
\includegraphics[clip=on,width=80mm,angle=0]{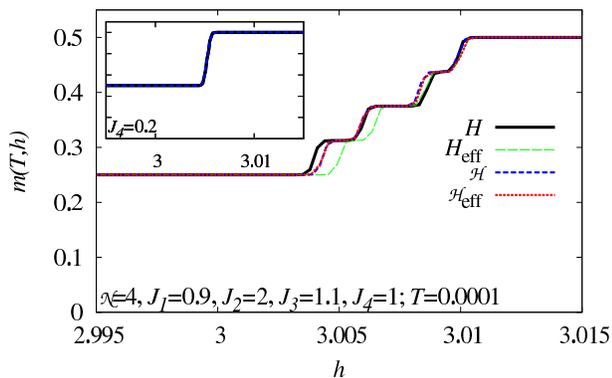}
\caption
{(Color online)  
Comparison of the full model with effective models for the distorted dimer-plaquette chain of ${\cal{N}}=4$ cells:  
Field dependences of the magnetization (per site) 
for $J_1=0.9$, $J_2=2$, $J_3=1.1$, $J_4=1$ (main panel) and $J_4=0.2$ (inset) at $T=0.0001$.
Note that for $J_4=0.2$ (inset) all curves practically coincide. 
For more explanations, see the main text.}
\label{fig04}
\end{center}
\end{figure}

\begin{figure}
\begin{center}
\includegraphics[clip=on,width=80mm,angle=0]{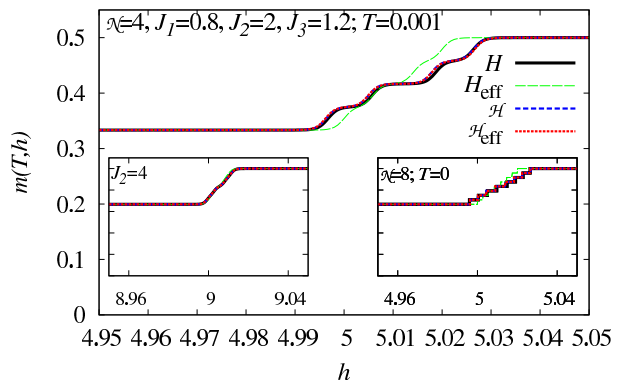}\\
\vspace{5mm}
\includegraphics[clip=on,width=80mm,angle=0]{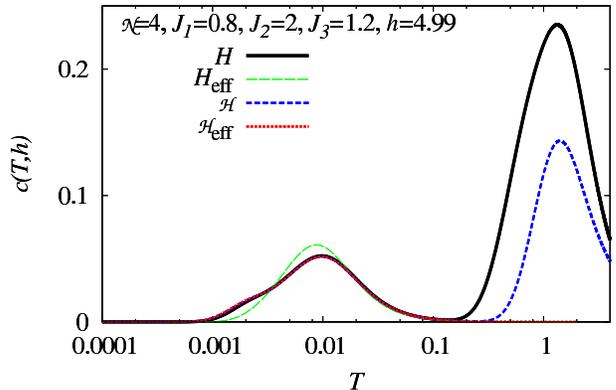}\\
\vspace{3mm}
\includegraphics[clip=on,width=80mm,angle=0]{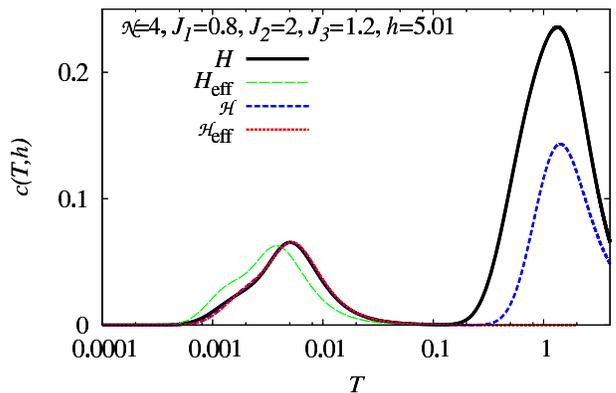}
\caption
{(Color online) 
Comparison of the full model with effective models for the distorted square-kagome lattice of ${\cal{N}}=4$ (main panel and left inset) and ${\cal{N}}=8$ (right inset): 
Field dependences of the magnetization (per site) and temperature dependences of the specific heat (per site) 
for $J_1=0.8$, $J_2=2$ (main panels and right inset) and $J_2=4$ (left inset), $J_3=1.2$.
For more explanations, see the main text.}
\label{fig05}
\end{center}
\end{figure}

In Figs.~\ref{fig03}, \ref{fig04}, and \ref{fig05},  
we show the magnetization curve at low (including zero) temperatures for the considered distorted models 
of 
${\cal{N}}=6$ (diamond chain,  Fig.~\ref{fig03}), 
${\cal{N}}=6$ (dimer-plaquette chain, Fig.~\ref{fig04}) 
as well as
${\cal{N}}=4$ and $8$ cells (square-kagome lattice, Fig.~\ref{fig05}).
In Figs.~\ref{fig03} and \ref{fig05},
we also show the temperature dependence of the specific heat at high fields.
Results for the initial full model (\ref{201}) 
(thick solid black curves)
are compared with corresponding data for the effective models 
[(i) strong-coupling approach, 
Hamiltonian $H_{\rm{eff}}$, 
see Eqs.~(\ref{304}) and (\ref{305}), (\ref{304}) and (\ref{306}), and (\ref{307}) and (\ref{308}), 
thin long-dashed green curves; 
(ii) localized-magnon description, 
Hamiltonian $\cal H$, 
see Eqs.~(\ref{310}), (\ref{315}), and (\ref{318}), 
short-dashed blue curves; 
(iii) localized-magnon description with subsequent perturbation approach, 
Hamiltonian ${\cal H}_{\rm{eff}}$, 
see Eqs.~(\ref{312}) and (\ref{313}), (\ref{312}) and (\ref{317}), and (\ref{319}) and (\ref{320}), 
dotted red curves].


Comparing the full initial model with the effective ones 
we find a quite excellent agreement with the localized-magnon based effective models described by $\cal H$ and ${\cal H}_{\rm eff}$. 
Note that in our plots the deviation from ideal geometry, i.e., $J_1=J_3$, is up to 20\%.
We mention that even for fairly large deviations, not shown here, the agreement is still satisfactorily.
Moreover, the perturbative localized-magnon description by ${\cal H}_{\rm eff}$ 
almost coincides with non-perturbative localized-magnon description by ${\cal H}$
for temperatures including the full low-temperature maximum in $c(T)$.  
Hence, the former one is the favored approach, since it contains much less degrees of freedom and therefore its treatment is simpler.
The neglected degrees of freedom become relevant only at higher temperatures.
On the other hand, 
for $J_2=3$ and $J_2=2$ considered in Figs.~\ref{fig03}, \ref{fig04}, and \ref{fig05}, the strong-coupling approach is significantly less accurate. 
Naturally, it becomes better by increasing of $J_2$, 
see, e.g., the inset in the upper panel of Fig.~\ref{fig03}. 
In general, we may conclude that the derived effective models approximate the exact results remarkably well for a wide range of parameters.

Some generic features emerging due to the deviation from the ideal geometry 
are already clearly visible from the data obtained for the initial full models of small size.
First we mention that the deviation from ideal geometry, i.e., $J_1 \ne J_3$,
does not yield a noticeable change of the width of the plateau 
which precedes the magnetization jump to saturation in the ground state 
(note that the full plateau is not shown in Figs.~\ref{fig03}, \ref{fig04}, and \ref{fig05}, 
where data near the upper end of this plateau are presented only).
However, 
the deviation from ideal geometry has a significant influence on the magnetization jump at $h_1$ present for  $J_1 = J_3$. 
Instead of the jump there is a (small) finite region,
$h_{1l}\le h \le h_{1h}$, 
where the low-temperature magnetization shows steep increase between two plateau values.
While the strong-coupling approach predicts the boundaries of this region $h_{1l}$ and $h_{1h}$ only qualitatively 
and underestimates the width of the region $h_{1h}-h_{1l}$,
the localized-magnon approach gives much better results for $h_{1l}$ and $h_{1h}$.
Further we notice that the specific heat $c(T)$ shows a two-peak temperature profile, 
typical for localized-magnon systems, also for the distorted models. 
The double-peak structure is even present for magnetic fields within the region  $h_{1l} <  h < h_{1h}$. 
All effective Hamiltonians are capable to reproduce the low-temperature peak of $c(T)$ 
related to an extra low-energy scale set by the manifold of almost localized-magnon states. 
Again the localized-magnon approach yields much better agreement with the full model than the strong-coupling approach.
Note that the effective model ${\cal{H}}$ can reproduce qualitatively even the second high-temperature maximum of the specific heat, 
see Figs.~\ref{fig03} and \ref{fig05}.
However, for the description of the low-temperature thermodynamics  the simpler effective model ${\cal{H}}_{\rm{eff}}$ is favorable, 
since both effective localized-magnon based models yield almost perfect agreement with the full model. 

Before we will discuss the low-temperature physics of the models at hand in more detail in the next section, 
let us summarize the main findings of the present section relevant for the further considerations:
We have presented three types of effective models 
which are capable to describe frustrated quantum antiferromagnets of the monomer universality class\cite{epjb2006} at high fields and low temperatures.
All effective models refer to a reduced Hilbert space (only two states for each trapping cell are taken into account),
the strong-coupling approach assumes $J/J_2$ to be small, 
whereas the localized-magnon approach needs a more modest requirement: The deviation from the ideal geometry should be small.
Comparisons of exact diagonalization data illustrate the quality of the suggested models.
The fairly simple and well investigated spin-1/2 $XXZ$ model with uniform nearest-neighbor interaction in a $z$-aligned magnetic field 
on a chain or on a square lattice, respectively, 
provides an accurate low-temperature description of the frustrated initial spin-1/2 Heisenberg models.
In the limit of sufficiently strong $J_2$ the even simpler spin-1/2 isotropic $XY$ model is adequate.

We may conclude the above considerations by the statement:
Instead of a classical hard-core models adequate to describe the low-energy degrees of freedom for ideal geometry 
we are faced with quantum models in the case of non-ideal geometry. 
However, these quantum models are well-known standard ones,
they are unfrustrated and they allow a straightforward application of efficient methods of quantum magnetism 
such as Bethe ansatz, density matrix renormalization group or quantum Monte Carlo techniques.
Moreover, for the effective models the particle number is smaller than that of the initial models.
{\it Finding of effective Hamiltonians is a key result of our paper}, from which several consequences follow to be discussed below.

\section{High-field low-temperature thermodynamics}
\label{sec4}
\setcounter{equation}{0}

Now we discuss in more detail the properties of the considered frustrated quantum antiferromagnets in the high-field low-temperature regime
using the effective models derived in the previous section.
For that we use the fact that the effective models, 
the spin-1/2 isotropic $XY$ or $XXZ$ Heisenberg model on a chain or on a square lattice,
were extensively studied by different analytical and numerical methods in the past.
Hence we can use this knowledge to discuss the physical properties of the much more complicated initial frustrated models in the relevant regime.

\subsection{The one-dimensional case}

\begin{figure}
\begin{center}
\includegraphics[clip=on,width=80mm,angle=0]{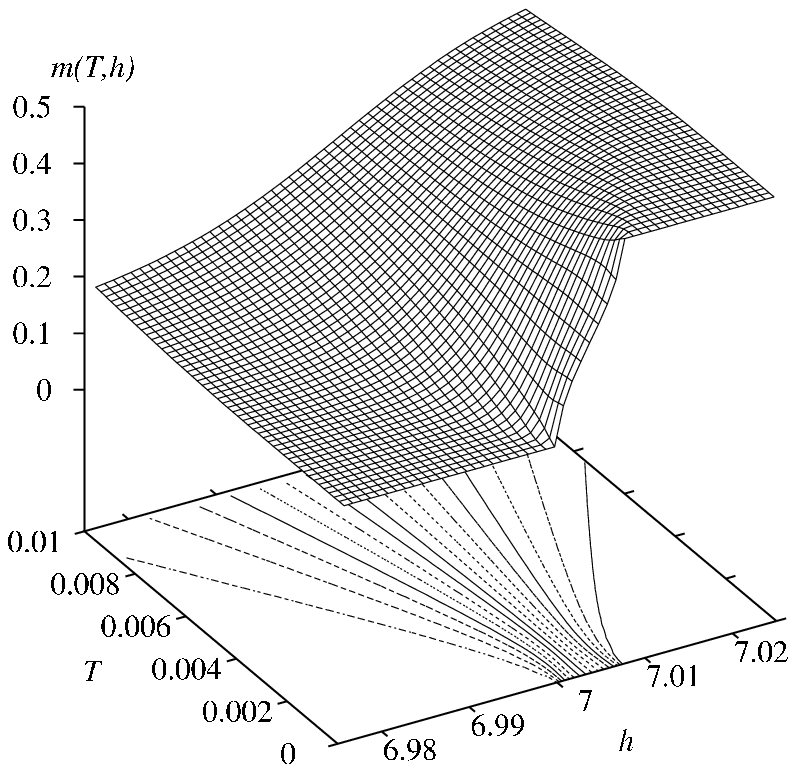}
\includegraphics[clip=on,width=80mm,angle=0]{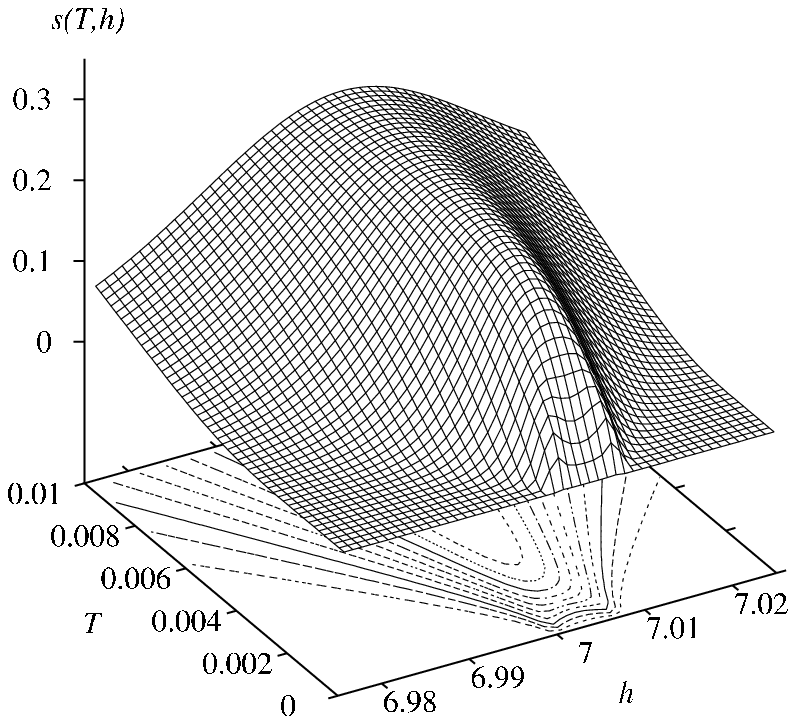}
\includegraphics[clip=on,width=80mm,angle=0]{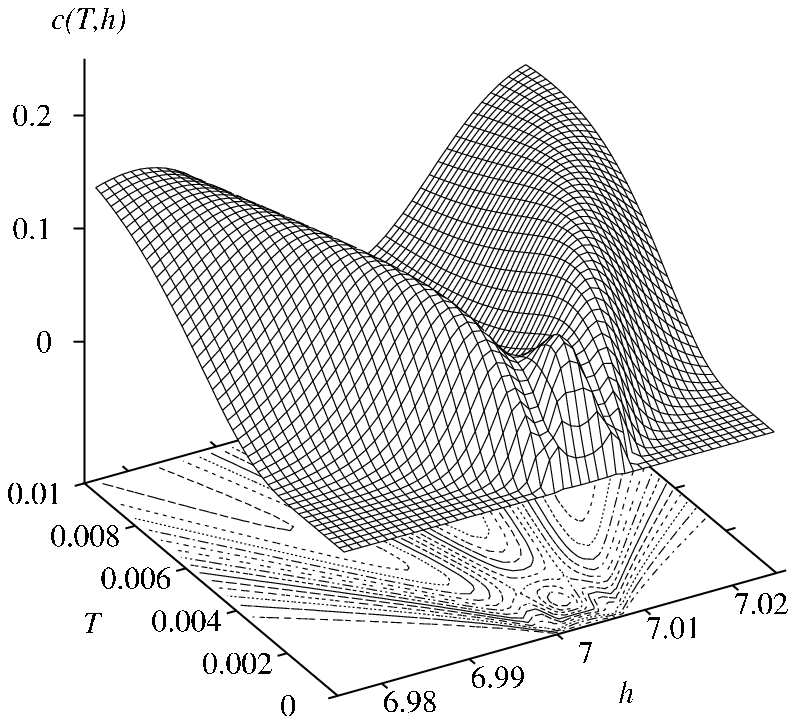}
\caption
{Magnetization per site $m(T,h)$, entropy per site $s(T,h)$, and specific heat per site $c(T,h)$ at high fields and low temperatures
for the distorted diamond-chain Heisenberg antiferromagnet (\ref{201}) with $J_1=0.85$, $J_2=6$, $J_3=1.15$
according to the effective Hamiltonian $H_{\rm{eff}}$ (\ref{304}), (\ref{305}).}
\label{fig06}
\end{center}
\end{figure}

The spin-1/2 isotropic $XY$ chain in a transverse field is a famous exactly solvable quantum model.\cite{lieb}
Therefore, 
the one-dimensional systems considered here
can be described analytically even in the thermodynamic limit $\cal{N}\to\infty$
if  the strong-coupling approach is adequate, i.e., if $J_2$ is sufficiently large.
In Fig.~\ref{fig06} we report some thermodynamic quantities 
for the distorted diamond-chain Heisenberg antiferromagnet (\ref{201}) with $J_1=0.85$, $J_2=6$, $J_3=1.15$
(cf. the inset in the upper panel of Fig.~\ref{fig03})
obtained by using the effective Hamiltonian $H_{\rm{eff}}$ given in Eqs.~(\ref{304}), (\ref{305}).

From the upper panel of Fig.~\ref{fig06} it is obvious 
that the ground-state magnetization jump at $h_1=7$ existing for ideal geometry, $J_1=J_3$,
transforms into the smooth magnetization curve of the spin-1/2 isotropic $XY$ chain in a transverse magnetic field for the distorted model. 
This smooth magnetization curve has an infinite slope with exponent 1/2 
when approaching continuously the plateaus at $h_{1l}=h_1=7$ and $h_{1h}=h_1+(J_1-J_3)^2/(2J_2)=7.0075$.
Furthermore, 
the finite residual ground-state entropy per site  at the saturation field, $s(T=0,h=h_1)=\ln(2)/3\approx 0.231$, present for  $J_1=J_3$, 
is removed immediately by distortions, i.e., $s(T=0,h)=0$. 
However, by a slight increase of the temperature 
the whole manifold of almost localized-magnon states becomes accessible for $h_{1l} \lesssim h \lesssim h_{1h}$, 
thus producing a tremendous entropy enhancement,
see the corresponding panel in Fig.~\ref{fig06}.
The behavior of the specific heat, shown in the lower panel of Fig.~\ref{fig06}, depends on the value of $h$.
For $h_{1l} \le h\le h_{1h}$ one has $c(T)\propto T$ for $T \to 0$ 
which corresponds to the gapless spin-liquid phase of the effective spin-1/2 isotropic $XY$ chain in a transverse magnetic field.
Otherwise there is an exponential decay of $c(T)$ for $T \to 0$ 
(within the gapped phase of the effective spin-1/2 isotropic $XY$ chain in a transverse field).
Correspondingly, the character of maxima in $c(T)$  seen in the lower panel of Fig.~\ref{fig06}
(representing the extra low-temperature maxima of the full model)  
is different. 
It is that one for the spin-1/2 isotropic $XY$ chain in the spin-liquid phase when $h_{1l}\le h\le h_{lh}$,
but it is the one for (weakly interacting) spins in a field if $h$ is significantly below (above) $h_{1l}$ ($h_{1h}$).
A crossover between these types of behavior produces interesting behavior of the position and the height of the low-temperature maximum of $c(T,h)$.

Qualitatively the behavior shown in Fig.~\ref{fig06} remains in the case of $XXZ$ (pseudo)spin interactions as well as in the two-dimensional case,
although quantitative details are obviously different
[cf., for example, Figs.~\ref{fig08a} and \ref{fig08b} which show results for the (effective) square-lattice $XXZ$ model
using a slightly different/complementary format of representation].
But, clearly, it is much harder to obtain analytical\cite{ba1,ba2} or numerical results for $XXZ$ models.

\subsection{Azurite}

Next, 
we apply the elaborated effective description to discuss some low-temperature properties 
of natural mineral azurite Cu$_3$(CO$_3$)$_2$(OH)$_2$,\cite{ewing1958}  
where the magnetization curve is experimentally accessible even beyond the saturation field.\cite{Kikuchi}
It is known that the magnetic properties of azurite can be explained using the diamond-chain Hamiltonian (\ref{201})
with the following set of parameters:
$J_1=15.51$K, $J_2=33$K,  $J_3=6.93$K
with $h=g\mu_{\rm{B}}{\sf{H}}$, $g=2.06$, $\mu_{\rm{B}}\approx 0.67171$K/T,
see Refs.~\onlinecite{effective_xy2} and \onlinecite{azurite-parameters}.
Note, however, that  also a small exchange interaction $J_m=4.62$K between the sites $m,3$ and $m+1,3$ [see Fig.~\ref{fig01}(a)] may be relevant.
Neglecting $J_m$ 
the exchange parameters of our effective Hamiltonian ${\cal H}_{\rm{eff}}$, 
see Eqs.~(\ref{312}) and (\ref{313}),
are ${\sf{J}}=0.845$K, ${\sf{J}}^z=0.287$K, 
and the effective magnetic field is ${\sf{h}}=(1.384{\sf{H}}-44.778)$K, 
where ${\sf{H}}$ is the experimentally applied magnetic field measured in teslas.
This effective  Hamiltonian is appropriate  to describe the low-temperature properties of azurite above 30T.
Note that in Ref.~\onlinecite{effective_xy2} slightly larger values of the effective parameters ${\sf{J}}$ and ${\sf{J}}^z$ were obtained 
(taking into account  $J_m=4.62$K) 
by analyzing the excitations above the one-third plateau state 
and 
by fitting to the values of the upper edge field of the one-third plateau and of the saturation field.

\begin{figure}
\begin{center}
\includegraphics[clip=on,width=80mm,angle=0]{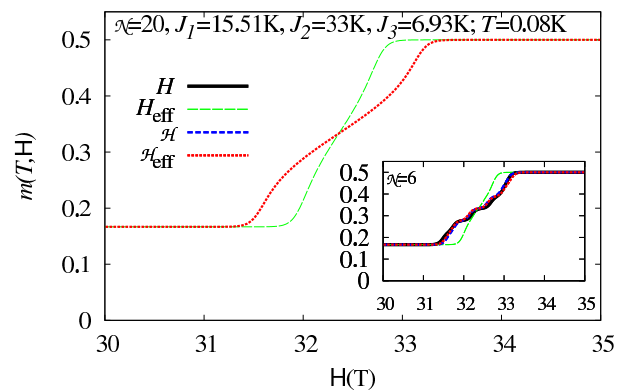}
\caption
{(Color online) 
Low-temperature ($T=0.08$K) magnetization curves for the distorted diamond-chain Heisenberg antiferromagnet (\ref{201}) 
with azurite parameters $J_1=15.51$K, $J_2=33$K,  $J_3=6.93$K ($J_m=0$), and the gyromagnetic ratio $g=2.06$.
Inset: ${\cal{N}}=6$, i.e., $N=18$; 
main panel:  ${\cal{N}}=20$, i.e., $N=60$.}
\label{fig07}
\end{center}
\end{figure}
\begin{figure}
\begin{center}
\includegraphics[clip=on,width=80mm,angle=0]{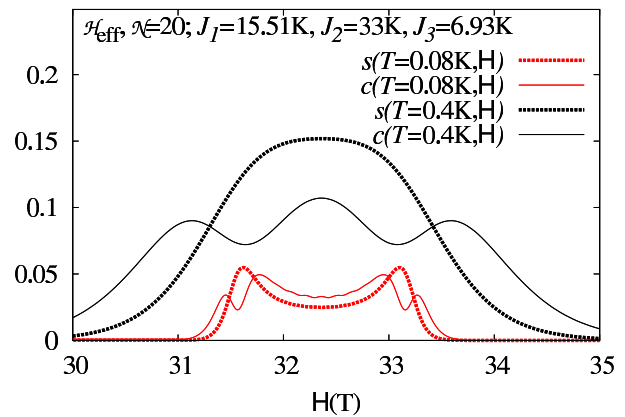}
\caption
{(Color online) 
Field dependence of the entropy per site (thick) and specific heat per site (thin) for the distorted diamond-chain Heisenberg antiferromagnet (\ref{201}) 
with azurite parameters $J_1=15.51$K, $J_2=33$K,  $J_3=6.93$K ($J_m=0$), and the gyromagnetic ratio $g=2.06$ 
at low temperatures, $T=0.08$K and $T=0.4$K.}
\label{fig07a}
\end{center}
\end{figure}
\begin{figure}
\begin{center}
\includegraphics[clip=on,width=80mm,angle=0]{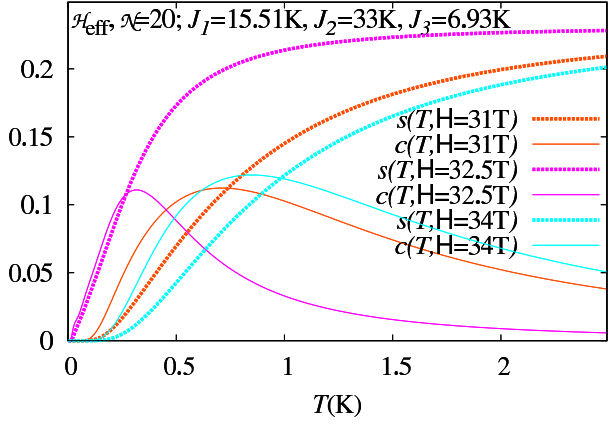}
\caption
{(Color online) 
Temperature dependence of the entropy per site (thick) and specific heat per site (thin) for the distorted diamond-chain Heisenberg antiferromagnet (\ref{201}) 
with azurite parameters $J_1=15.51$K, $J_2=33$K,  $J_3=6.93$K ($J_m=0$), and the gyromagnetic ratio $g=2.06$ 
at high magnetic fields, ${\sf{H}}=31$T, $32.5$T, and $34$T.}
\label{fig07b}
\end{center}
\end{figure}

We present  results for the azurite parameters  for various  temperatures, 
including  $T=0.08$K related to the measured magnetization curve,\cite{Kikuchi}
in Figs.~\ref{fig07}, \ref{fig07a}, and \ref{fig07b}.
In the inset of Fig.~\ref{fig07} we first compare the results  of different approaches for ${\cal{N}}=6$ (i.e., $N=18$). 
Again we observe that the results for $H$, ${\cal{H}}$, and  ${\cal{H}}_{\rm{eff}}$ agree very well,
whereas the data for the strong-coupling approach, $H_{\rm{eff}}$, deviate noticeably.
Hence we further focus on data obtained from  our effective Hamiltonian ${\cal{H}}_{\rm{eff}}$ (\ref{312}), (\ref{313}) for ${\cal{N}}=20$, i.e., $N=60$, 
with the parameters ${\sf{J}}=0.845$K, ${\sf{J}}^z=0.287$K, and ${\sf{h}}=(1.384{\sf{H}}-44.778)$K.
The magnetization curve at $T=0.08$K is shown in Fig.~\ref{fig07},
the field dependences of the specific heat per site $c$ and the entropy per site $s$ at $T=0.08$K and $T=0.4$K are shown in Fig.~\ref{fig07a},
and in Fig.~\ref{fig07b} we present the temperature dependences of the specific heat $c$ and the entropy $s$
for three values of the  magnetic field, ${\sf{H}}=31$T, $32.5$T, and $34$T.
Remember that the effective model ${\cal{H}}_{\rm{eff}}$ covers the low-temperature region, 
whereas the high-temperature maximum in $c(T)$ and the corresponding entropy $s \to \ln 2 \approx 0.693$ are not provided by ${\cal{H}}_{\rm{eff}}$.

There are well pronounced features below 1K which can be attributed to traces of independent localized-magnon states:
A steep increase of the magnetization (Fig.~\ref{fig07}),
an enhanced low-temperature entropy that reaches $\ln(2)/3 \approx 0.231$ (Fig.~\ref{fig07b}),
and
a low-temperature maximum in the specific heat (Fig.~\ref{fig07b}).
The large entropy change caused by variation of the magnetic field (see Fig.~\ref{fig07a}) implies an enhanced magnetocaloric effect
as it was noticed already in Refs.~\onlinecite{zhito_hon2004} and \onlinecite{effective_xy2}.
[Note that recently the magnetocaloric effect in an $XXZ$ chain has been examined rigorously.\cite{trippe}]
It is in order to mention here recent experiments on the magnetocaloric effect
for other (frustrated) quantum Heisenberg antiferromagnets.\cite{mce_exp}
Corresponding experimental magnetocaloric studies for azurite would be of great interest.

\subsection{The two-dimensional case}

The effective Hamiltonians for the distorted square-kagome lattice are well-known spin-1/2 square-lattice models:
The ${\cal{N}}$-site square-lattice isotropic $XY$ model (\ref{307}),
the $3{\cal{N}}$-site $XXZ$ model on a decorated square lattice (\ref{318}),
and
the ${\cal{N}}$-site square-lattice $XXZ$ model (\ref{319}).
All models contain a Zeeman term with a magnetic field in $z$ direction.  
The first and the third models have been studied quite extensively in the past
(also in the context of hard-core bosons;
then the magnetization corresponds to the particle number and the magnetic field to
the chemical potential).\cite{qmc_d1,qmc_d2,domanski,harada,qmc_s,tognetti,hcbosons_1,hcbosons_2}

It is generally known that the classical version of the isotropic $XY$ model (without field) 
undergoes a Berezinskii-Kosterlitz-Thouless (BKT) transition\cite{bkt} at $T_c\approx 0.898 \vert{\sf{J}}\vert$.\cite{gupta}
The BKT transition occurs in the quantum spin-1/2 case too, although, at a lower temperature, $T_c\approx 0.343 \vert{\sf{J}}\vert$.\cite{qmc_d1,harada,tognetti}
The quantum model is gapless with an excitation spectrum that is linear in the momentum.
The specific heat shows $T^2$ behavior for $T \to 0$, 
it increases very rapidly around $T_c$, 
and it exhibits a finite peak somewhat above $T_c$.
This kind of the low-temperature thermodynamics survives for not too large $z$-aligned magnetic field $\vert {\sf{h}}\vert < 2\vert {\sf{J}}\vert$
(see,  e.g., Figs.~3 and 8 in Ref.~\onlinecite{hcbosons_2}).
Also for the spin-1/2 square-lattice $XXZ$ model with dominating isotropic $XY$ interaction in a $z$-aligned  magnetic field the BKT transition appears.\cite{hcbosons_1}

\begin{figure}
\begin{center}
\includegraphics[clip=on,width=80mm,angle=0]{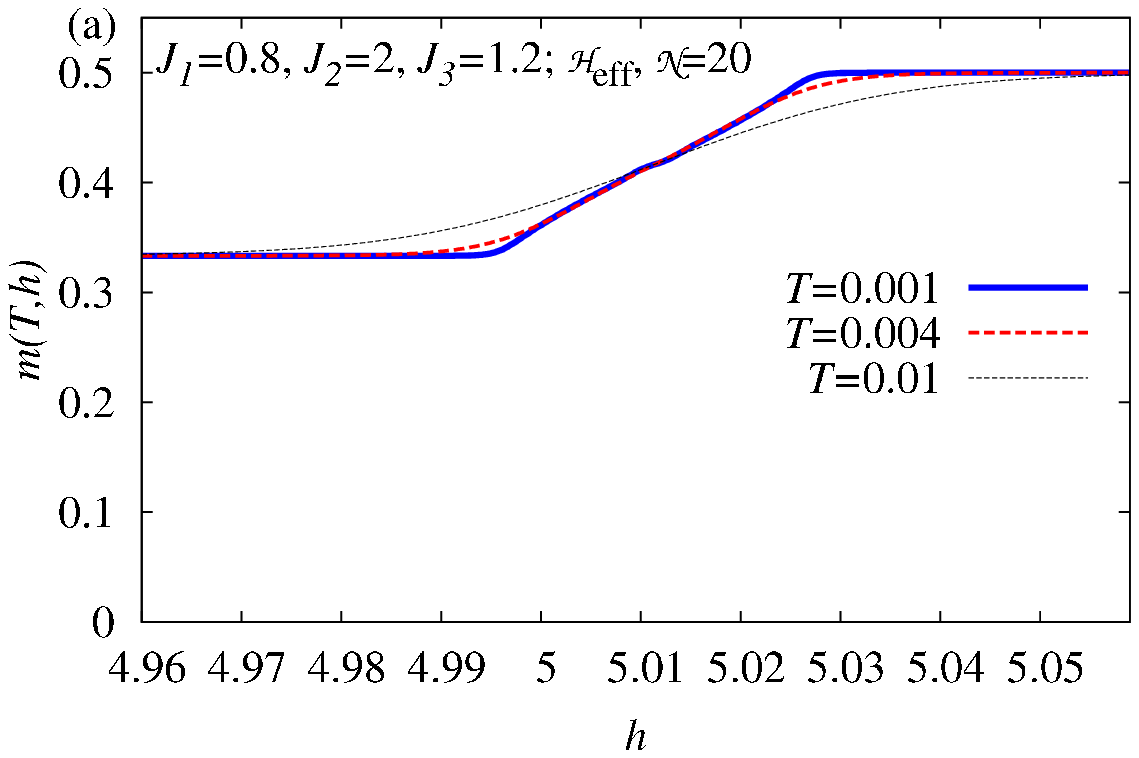}\\
\vspace{3mm}
\includegraphics[clip=on,width=80mm,angle=0]{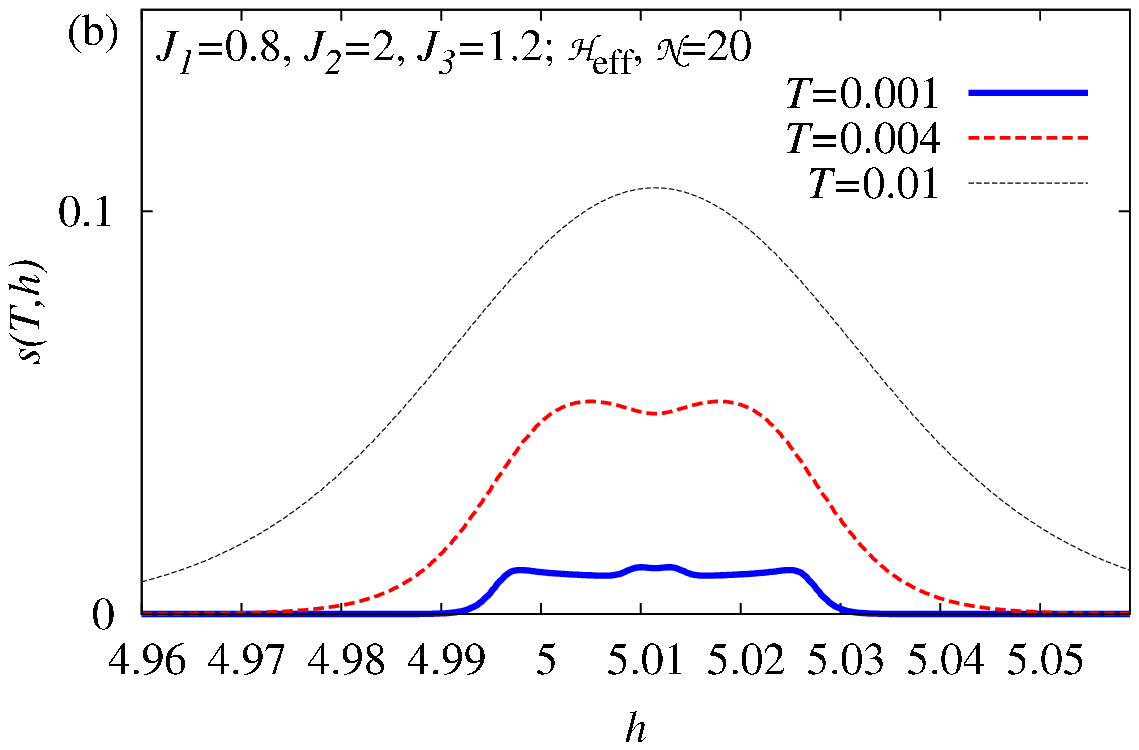}\\
\vspace{3mm}
\includegraphics[clip=on,width=80mm,angle=0]{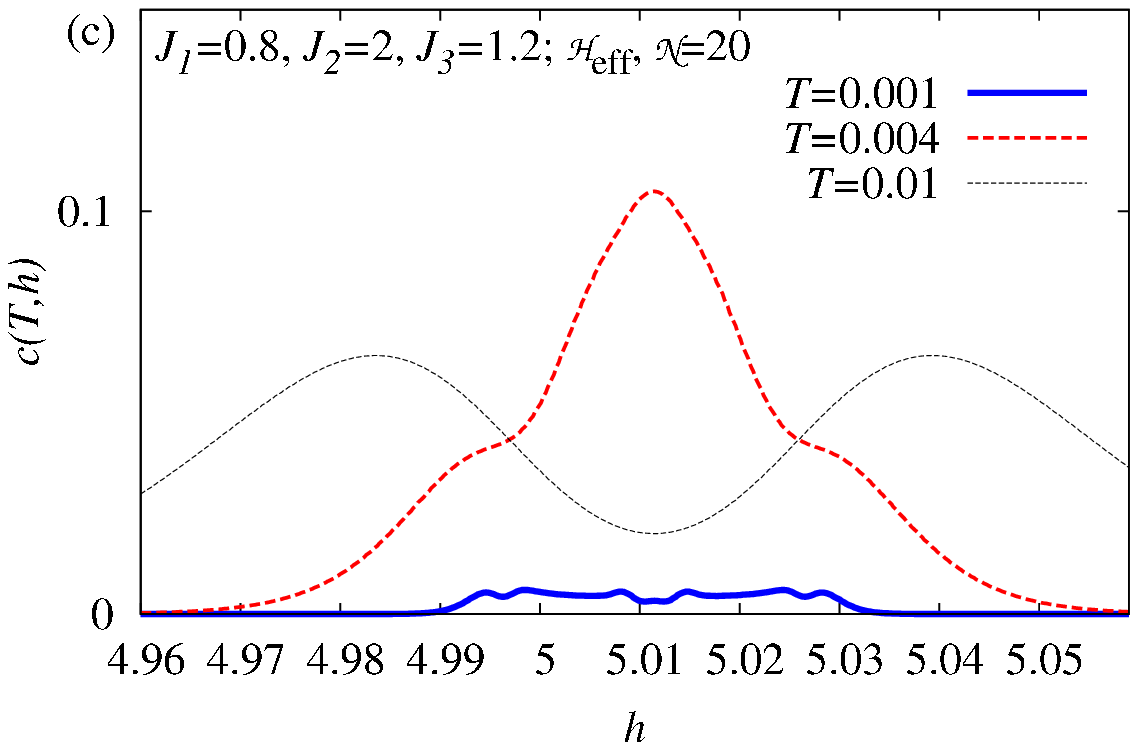}
\caption
{(Color online) 
(a) Magnetization per site $m(T,h)$, 
(b) entropy per site $s(T,h)$, 
and 
(c) specific heat per site $c(T,h)$ 
at high fields and low temperatures ($T=0.001,0.004,0.01$)
for the distorted square-kagome Heisenberg antiferromagnet (\ref{201}) with $J_1=0.8$, $J_2=2$, $J_3=1.2$.
The data are obtained from the effective Hamiltonian ${\cal{H}}_{\rm{eff}}$ (\ref{319}), (\ref{320}) with ${\cal{N}}=20$.}
\label{fig08a}
\end{center}
\end{figure}
\begin{figure}
\begin{center}
\includegraphics[clip=on,width=80mm,angle=0]{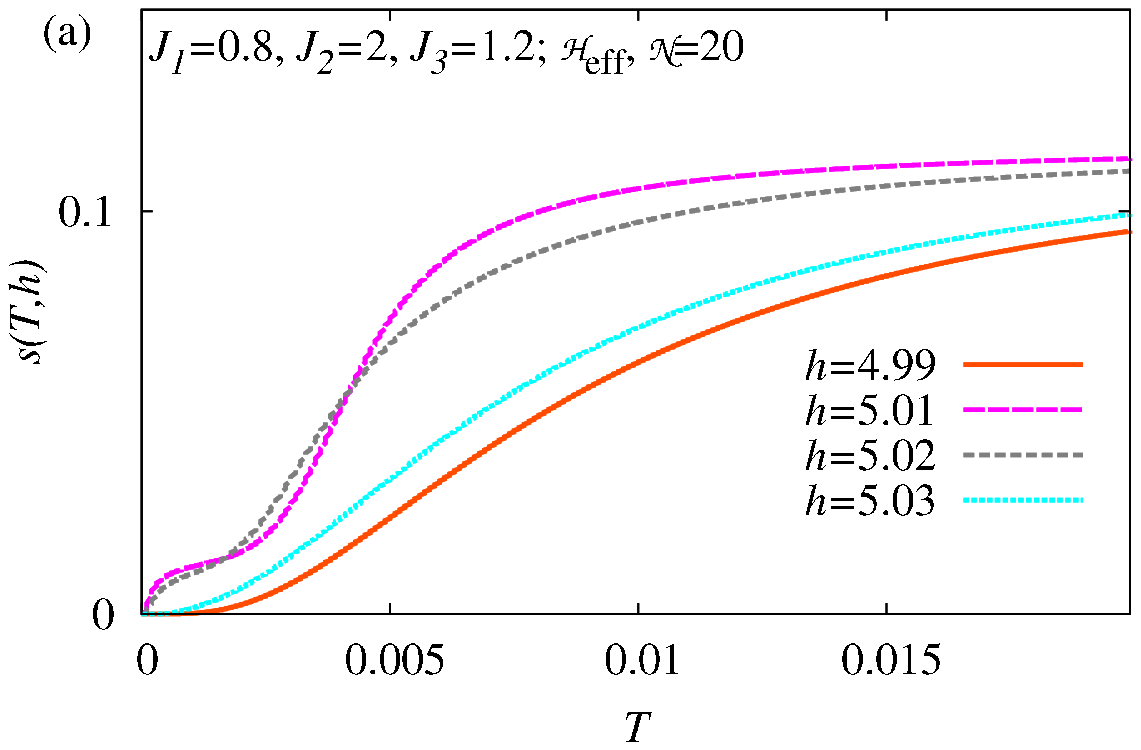}\\
\vspace{3mm}
\includegraphics[clip=on,width=80mm,angle=0]{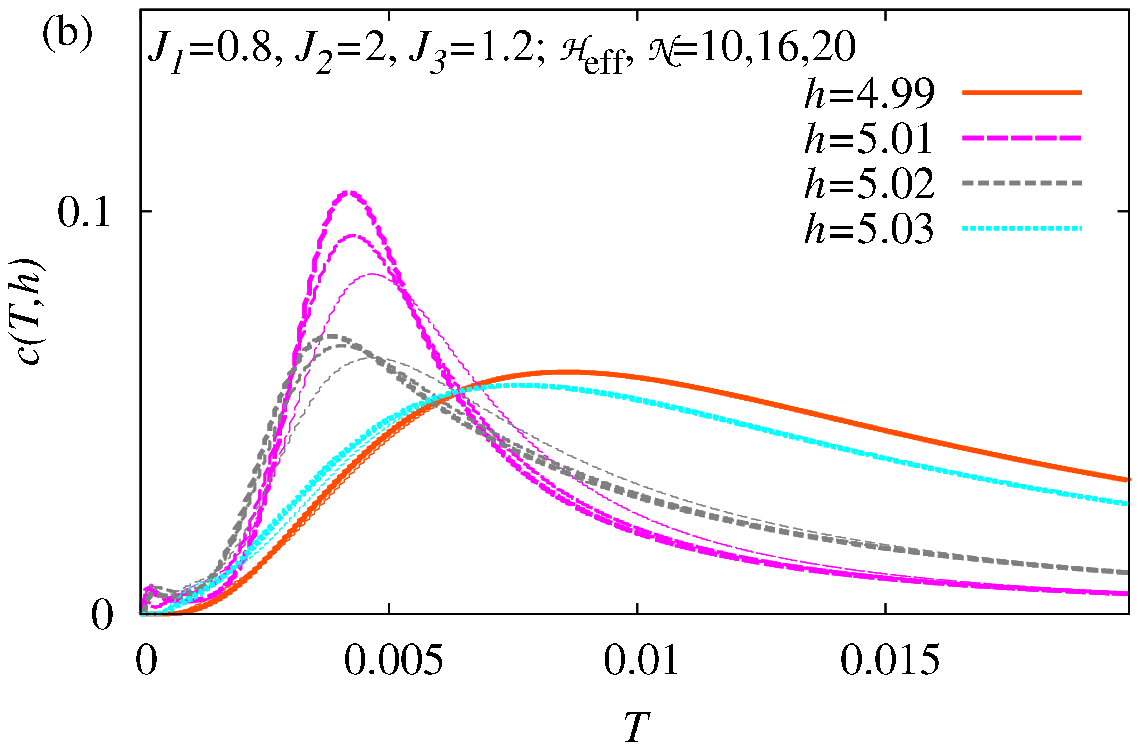} 
\caption
{(Color online) 
(a) Entropy per site $s(T,h)$ 
and 
(b) specific heat per site $c(T,h)$ 
at high fields ($h=4.99,5.01,5.02,5.03$) and low temperatures
for the distorted square-kagome Heisenberg antiferromagnet (\ref{201}) with $J_1=0.8$, $J_2=2$, $J_3=1.2$
according to exact diagonalization data for the effective Hamiltonian ${\cal{H}}_{\rm{eff}}$ (\ref{319}), (\ref{320}) with ${\cal{N}}=20$.
Temperature dependence of $c(T,h)$ are shown also for ${\cal{N}}=16$ (thin curves) and ${\cal{N}}=10$ (very thin curves).}
\label{fig08b}
\end{center}
\end{figure}

These known results can be translated to the considered case,
i.e., the distorted square-kagome spin model (\ref{201}) in the high-field low-temperature regime.
Again we use exact diagonalization for the effective model ${\cal{H}}_{\rm{eff}}$ (\ref{319}), (\ref{320}) of ${\cal{N}}=20$ sites
for the set of parameters $J_1=0.8$, $J_2=2$, $J_3=1.2$ to analyze the low-temperature features of the initial $N$-site square-kagome model 
(remember that ${\cal{N}}=20$ corresponds to $N=120$). 
From the previous section we know that ${\cal{H}}_{\rm{eff}}$ for this set of parameters works perfectly well at least up to $T=0.1$,
cf. Fig.~\ref{fig05}. 
Our results are collected in Figs.~\ref{fig08a} and \ref{fig08b}.
Instead of the former jump of the ground-state magnetization per site between 1/3 and 1/2 present at $h_1=5$ for $J_1=J_3=1$,
the magnetization $m(T,h)$ shows a smooth (but steep) increase varying the field $h$ from $h_{1l}\approx 4.996$ to $h_{1h}\approx 5.027$, 
see Fig.~\ref{fig08a}(a).
Due to the distortion, 
the former $2^{{\cal{N}}}$-fold  degeneracy of the ground states at saturation field is lifted 
and, as a result, there is no residual entropy at $T=0$.
Since these $2^{{\cal{N}}}$ energy levels remain close to each other, 
in the field region $h_{1l}\le h\le h_{1h}$  by a slight increase of $T$ all these states become accessible 
and the entropy shows clear traces of the former residual entropy of size $\ln(2)/6\approx 0.116$, 
see  Figs.~\ref{fig08a}(b) and \ref{fig08b}(a) and cf. also Ref.~\onlinecite{localized_magnons2}.
Concerning the behavior of the specific heat of the distorted square-kagome model for $T\to 0$ 
we use the knowledge for the energy spectrum of the effective easy-plane $XXZ$ model: 
The spectrum is gapless with a linear dispersion of excitations in the field region $h_{1l}\le h\le h_{1h}$
and, as a result, the specific heat is $\propto T^2$ at low temperatures.
Outside that field region the  spectrum is gapped leading to an exponential decay of the specific heat for $T\to 0$.

\begin{figure}
\begin{center}
\includegraphics[clip=on,width=80mm,angle=0]{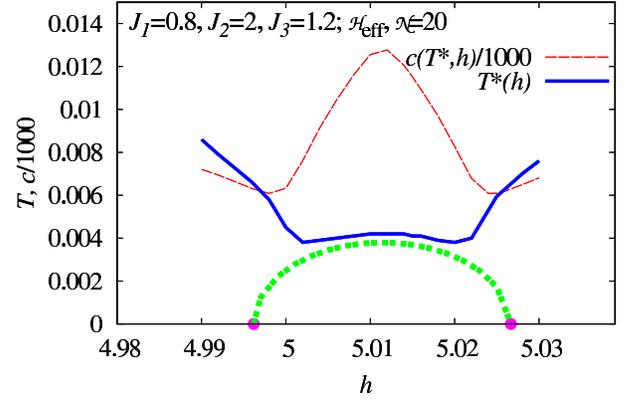}
\caption
{(Color online) 
Sketch of the phase diagram of the distorted square-kagome Heisenberg antiferromagnet 
($J_1=0.8$, $J_2=2$, $J_3=1.2$) 
at high magnetic field  (thick dashed green line)  
as it is indicated by the position of the maximum $T^{*}$ of the specific heat $c(T,h)$ (thick solid blue curve).
Thin dashed red curve corresponds to the value of $c(T^{*},h)$ divided by 1000.
By filled violet circles the values of $h_{1l}$ and $h_{1h}$ are indicated.}
\label{fig09}
\end{center}
\end{figure}

With respect to the BKT transition the low-temperature maximum of the specific heat, 
cf. Figs.~\ref{fig08a}(c) and \ref{fig08b}(b), 
deserves a more detailed discussion.
For the gapped spectrum there is a broader maximum in $c(T)$.
By contrast, in the gapless field region $h_{1l}\le h\le h_{1h}$, the maximum occurs at lower temperatures and it is more pronounced, 
i.e., it becomes peak-like
[compare, e.g., the curves for $h=4.99$ and $h=5.01$ in Fig.~\ref{fig08b}(b)].
As mentioned above, this well-pronounced maximum is located somewhat above the BKT transition point $T_c$. 

Comparing the data for ${\cal{N}}=10$, 16, and 20, see Fig.~\ref{fig08b}(b), 
one clearly  see that the maximum in the $c(T)$ curves corresponding to $h_{1l}<h<h_{1h}$ 
shows significant finite-size effects (in particular the height of the maximum increases noticeably with system size ${\cal{N}}$),
whereas in the gapped regime the  $c(T)$ curves are insensitive to the system size also around the maximum.
The size dependence of the height of the maximum can be interpreted as signature of the BKT transition 
(see also the discussion in Ref.~\onlinecite{qmc_d1}).

Using the maximum in the specific heat as an indicator of the BKT transition in the distorted square-kagome Heisenberg antiferromagnet 
we can construct a sketch of the phase diagram of the model in the  $h$ -- $T$ plane, see Fig.~\ref{fig09}.
The largest transition temperature $T_c$ appears for zero effective field ${\sf{h}}$, 
which corresponds to $h \approx 5.011$, of the initial model, see Eq.~(\ref{320}).
A finite effective field $|{\sf{h}}|>0$ yields a decrease in $T_c$, 
finally $T_c$ becomes zero entering the gapped phase at $h_{1l}$ or at $h_{1h}$.
As mentioned above, these features appear only at very low temperatures.
By lowering  of $J_2$ and/or increasing of $|J_1-J_3|$ one can get effective exchange parameters being larger by one order of magnitude, 
see Eq.~(\ref{320}), 
and correspondingly enlarged $T_c$.
Discussing the field dependence of $T_c$, sketched in Fig.~\ref{fig09}, in terms of the initial model
we may conclude 
that in the highly frustrated quantum Heisenberg antiferromagnet on the square-kagome lattice with deviations from ideal geometry 
the BKT transition appears only for non-zero magnetic fields, 
i.e., we are faced with a magnetic-field driven BKT transition.

\section{Conclusions}
\label{sec5}
\setcounter{equation}{0}

Motivated by recent experiments on frustrated quantum Heisenberg antiferromagnets and recent theories of localized-magnon systems
we have investigated the high-field low-temperature regime of various frustrated quantum spin systems 
belonging to the monomer class of localized-magnon systems.
In the case of the ideal geometry, i.e., when the localization conditions are strictly fulfilled, 
their low-temperature physics is well described by noninteracting (pseudo)spins 1/2 in a field. 
This description  is equivalent to the mapping onto hard-core objects used in previous papers. 
Deviations from ideal geometry, 
which are relevant for possible experimental detection of the features related to localized magnons, 
set a new low-energy scale determined  by the distortion of the exchange constants.
An effective description of the corresponding low-energy physics for the distorted systems is given by non-frustrated $XXZ$ models, 
i.e., collective quantum phenomena emerge. 
New generic features due to the deviation form ideal geometry appear 
for magnetic fields in the vicinity of the former saturation field $h_1$ of the undistorted systems, 
namely:
(i) the ground-state magnetization curve exhibits a smooth (but steep) part 
(instead of perfect jump at $h_1$),
(ii) there is a drastic enhancement of the entropy at very small but nonzero temperatures 
(instead of nonzero residual ground-state entropy at $h_1$),
(iii) the specific heat exhibits a power-law decay 
(instead of zero specific heat at $h_1$ and exponentially vanishing specific heat just in the vicinity of $h_1$).
It is worth noting that strong variation of entropy with varying magnetic field 
leads to a noticeable magnetocaloric effect at low temperatures around the saturation field.
Moreover, we mention that the characteristic extra low-temperature peak in the specific heat survives in distorted systems.  
The most prominent collective phenomenon is the appearance of a BKT transition driven by the magnetic field 
in the considered two-dimensional distorted localized-magnon system, the square-kagome Heisenberg antiferromagnet. 
Likely, such a scenario is not restricted to this particular model,
rather it should be also present in other two-dimensional quantum Heisenberg systems built by weekly coupled localized-magnon cells.
We may conclude that the findings of our paper can be useful searching for experimental manifestation of localized-magnon effects.

\section*{Acknowledgments}

The numerical calculations were performed using J.~Schulenburg's {\it spinpack}.\cite{spinpack} 
The present study was supported by the DFG (project RI615/21-1).
O.~D. acknowledges the kind hospitality of the University of Magdeburg in October-December of 2012 and in March-May of 2013.
O.~D. would like to thank the Abdus Salam International Centre for Theoretical Physics (Trieste, Italy) 
for partial support of these studies through the Senior Associate award.
O.~D. and J.~R. are grateful to the MPIPKS (Dresden) 
for the hospitality during the International Focus Workshop ``Flat Bands: Design, Topology, and Correlations'' (6 - 9 March 2013).

\end{document}